\documentclass[reqno]{amsart}
\usepackage{amssymb}
\usepackage{lscape}
\usepackage{longtable}
\usepackage{amssymb,epsfig,graphics}
\usepackage{amsmath}
\usepackage{amscd}
\usepackage{verbatim}
\usepackage{abstract}
\allowdisplaybreaks[4]
\input xypic
\usepackage{subfigure}
\usepackage[toc]{appendix}

\newtheorem{prop}{Proposition}
\newtheorem{example}{Example}
\newtheorem{pf}{Proof}

\def\C{\mathbb C}
\def\R{\mathbb R}
\def\Z{\mathbb Z}

\def\r{\rangle}
\def\l{\langle}

\def\a{\alpha}

\def\l{\langle}
\def\r{\rangle}

\renewcommand{\maketitle}{

\begin{flushright}
 \begin{center}
{\Large\bf Complete Decompositions of \\Coxeter groups orbit products of $A_2$, $ C_2$, $G_2$ and $H_2$ and some limits for them}
\end{center}
\end{flushright}
\vspace{0.5cm}
\begin{center}
{\bf Agnieszka Tereszkiewicz}
\end{center}
\vspace{0.5cm}

\begin{center}
Institute of Mathematics,
University of Bialystok, Akademicka~2, PL-15-267,
Bialystok, Poland\\

\vspace{0.5cm}

Centre De Recherches Math\'ematiques, Universit\'e de Montr\'eal, C.~P.~6128, succ. Centre-ville, Montr\'eal, H3C\,3J7, Qu\'ebec, Canada\\
\textit{mail: a.tereszkiewicz@uwb.edu.pl}

\end{center}

\date{\today}
}

\begin{document}

\maketitle

\vspace{1cm}

 \begin{abstract}\

Orbits of the Weyl reflection groups attached to the simple Lie groups $ A_2 $, $ C_2 $, $ G_2 $  and Coxeter group $H_2$ are considered. For each of the groups products of any two orbits are decomposed into the union of the orbits. Results are presented in a single formula for each of the groups. Orbits are considered as functions of two variables and limits  for such functions are mentioned.
\end{abstract}
\vspace{0.5cm}

%%%%%%%%%%%%%%%%%%%%%%%%%%%%%%%%%%%%%%%%%%%%%%%%%%%%%%%%%%%%%%%
\section{Introduction}
%%%%%%%%%%%%%%%%%%%%%%%%%%%%%%%%%%%%%%%%%%%%%%%%%%%%%%%%%%%%%%

Finite reflections groups, called the Coxeter groups, split into two classes: crystallographic (Weyl groups of semisimple Lie algebras) and  \linebreak non-crystallographic groups \cite{H,K}. There are groups generated by $n$ reflections  which act in real $n$ dimensional euclidian space $\R^n$. All the distinct points that can be obtained by the action of $G$ on a selected single point (the `seed' point) of $\R^n$ form the Coxeter group orbit. %For our problem we choose the seed point to belong to weight lattice of the Lie algebra.
Every lattice point belongs to precisely one $G$-orbit.
Group orbits played very important role, specially in weight systems of finite dimensional irreducible representations of simple Lie algebras (crystallographic case). Each weight system is a union of several Weyl group orbits. To determine these elements for a given representation of a given simple Lie algebra, it is a computational problem for which an efficient algorithm is known \cite{MP82,BMP}. Unlike the weight systems  grow without limits with increasing representations and the $G$-orbits are finite in size for a given Lie algebra. Indeed, the largest number of distinct points (`weights') an orbit can have, is the order of the Weyl group of the Lie algebra.

The points that form a given orbit are relatively easy to calculate, starting from any one of them, by repeated application of the reflections generating the  Coxeter group $G$. One almost always chooses as the seed point for such a calculation the unique (`dominant') point of the orbit. This point is easily identified because it is the only one in every orbit that has non-negative coordinates in $\omega$-basis. A $G$-orbit of Coxeter group of rank $n$  can be viewed as an $n$-dimensional polytope \cite{HLP08} with orbit points as its vertices (0-dimensional faces). Faces of  dimensions up to $(n-1)$ are also readily described \cite{CKPS}.

Tensor products of irreducible representations of the Lie algebras are in one-to-one correspondence with the tensor product of the weight systems. Hence  the decomposition of product  into irreducible components leads naturally to the problem of decomposition of tensor products of Coxeter group orbits.

 Decomposition of products of weight systems into irreducible ones is a familiar problem in representation theory. It is  frequently calculated in terms of the products of weight systems. Complexity of the decomposition problem rapidly increases with increasing representation that are being multiplied. The only known way how to provide the decomposition of all products of representations for an algebra of rank greater than 1, is by means of the appropriate generating function \cite{PS1979}. Unfortunately such generating function is known only for  $ A_2 $.  Due to frequent use in  particle, nuclear and atomic physics, many special cases of product decompositions are found in the physics literature \cite{Sl,Ram}.

Computing separately decomposition of products of Coxeter group orbits allows one to simplify the task of  decomposing  of the products of the weights systems.

There are two computational task involved when one decomposes the product of weight system: \newline (i) Computing the multiplicity of the Coxeter group orbits in any representation involved.\newline
(ii) Decomposing products of the individual $G$-orbits. This is the problem solved in this paper.

The situation is quite different when one considers separately the second problem, namely decomposition of products of $G$-orbits into the union of individual orbits. Since the number of points in any orbit cannot exceed the order of the corresponding Coxeter group and it is easily determined in all cases, the decomposition problem for orbits is a finite one, no matter how large the dominant weights may be. Therefore at least, in principle, all the decompositions for a fixed group can be explicitly solved. For the first problem an  independent algorithmic solution is known \cite{MP82,BMP,Br}.

In the paper the second problem, decomposition of products of two orbits, is solved explicitly for the three crystallographic groups $ A_2 $,  $ C_2 $, $ G_2 $ and non-crystallographic group $H_2$. The solution is presented separately for each group and the sketch prove for $A_2$ is given. Demonstrations of formulae for  the others groups one could try to find analogously to the presented one. For the
case $A_2$ formula for the decomposition of product of two orbits was presented firstly in \cite{Ag}.

There are other problems where the $G$-orbits are essential. Description of reflection generated polytopes which in a simple version (all vertices belong to one $G$-orbit) is found in \cite{CKPS}, symmetries of Clebsch-Gordan coefficients for groups of rank $\geq2$ can be formulated in terms of the $G$-orbits \cite{MP84}. The $G$-orbits have been used in description of viruses \cite{T}.

Most predictable exploitation of Weyl group orbits one can used in extensive computations with representations of semisimple Lie algebras/groups such as decomposition of tensor products, see one of the largest examples in \cite{GP},  in branching rules computation, i.e. restriction to representation of subgroups, \cite{LNP,LP} or for the others \cite{LPS}.
Another exploitation for Coxeter group is in Fourier expansions of digital data on multidimensional lattices. Particularly when a series of similar size orbits will be needed.

Problem of decomposition of product of two orbits applies for finding product of two orbit functions as a sum of orbit functions of  the same type, corresponding to the same Coxeter group, see \cite{KP06}. Moreover formulas for decomposition of two orbits  work also in the case when the dominant points are points which coordinates are nonnegative real numbers. Orthogonality for nonnegative integers is know, see for example \cite{MP06}. It means that one could investigate orthogonality of  functions for which dominant points are nonnegative real numbers.

The orbit $O(\lambda)$ of the group $G$ can be considered also as a continues  function of two variables. Then limits for them  and for their product could be calculated. In the paper this problem is briefly described and some limits are presented.

%%%%%%%%%%%%%%%%%%%%%%%%%%%%%%%%%%%%%%%%%%%%%%%%%%%%%%%%%%%%%
\section{Preliminaries}

The Coxeter groups $ A_2 $, $ C_2$, $G_2$ and $H_2$ are groups of order $6, 8, 12$ and $10$, respectively,   generated by reflection in two mirrors intersecting under the angle $\frac{\pi}{3},\frac{\pi}{4},\frac{\pi}{6}$ and $\frac{\pi}{5}$ at the origin of the real Euclidean space $\R^2$. In physics these are the  dihedral groups of order $6,8,12,10$, respectively.

In order to treat all these cases in uniform way, it is advantageous  to work in $\R^2$ with a pair of dual bases. The $\alpha$-basis (simple root basis) is  defined by the scalar products
\begin{alignat*}{2}
A_2 &:\quad\l\alpha_1\mid\alpha_1\r=
\l\alpha_2\mid\alpha_2\r=2\,,\quad
&&\l\alpha_1\mid\alpha_2\r=-1,\\
C_2 &:\quad\l\alpha_1\mid\alpha_1\r=1\,,\
\l\alpha_2\mid\alpha_2\r=2\,,\quad
&&\l\alpha_1\mid\alpha_2\r=-1,\\
G_2  &:\quad\l\alpha_1\mid\alpha_1\r=2\,,\
\l\alpha_2\mid\alpha_2\r=\tfrac23\,,\quad
&&\l\alpha_1\mid\alpha_2\r=-1,\\
H_2&:\quad\l\alpha_1\mid\alpha_1\r=\l\alpha_2\mid\alpha_2\r=2\,,\quad
&&\l\alpha_1\mid\alpha_2\r=-\tau,\;\;\;\textrm{where }\,\tau=\tfrac12(1+\sqrt{5}).
\end{alignat*}
The $\omega$-basis is defined as dual to $\alpha$-basis,
\begin{equation}\label{weightreflect}
\langle\omega_k\mid\alpha_j\rangle
 =\frac{\l\a_j\mid\a_j\r}{2}\delta_{jk}.
 \end{equation}
  Explicitly,
\begin{equation*}
\begin{alignedat}{4}\label{bases}
A_2:&\;\;\omega_1=\tfrac23\alpha_1+\tfrac13\alpha_2\,,\ \
&&\omega_2=\tfrac13\alpha_1+\tfrac23\alpha_2\,,\ \
&&\alpha_1=2\omega_1-\omega_2\,,\ \
&&\alpha_2=-\omega_1+2\omega_2,\\
C_2:&\;\; \omega_1=\alpha_1+\tfrac12\alpha_2\,,\ \
&&\omega_2=\alpha_1+\alpha_2\,,\ \
&&\alpha_1=2\omega_1-\omega_2\,,\ \
&&\alpha_2=-2\omega_1+2\omega_2,\\
G_2:&\;\; \omega_1=2\alpha_1+3\alpha_2\,,\ \
&&\omega_2=\alpha_1+2\alpha_2\,,\ \
&&\alpha_1=2\omega_1-3\omega_2\,,\ \
&&\alpha_2=-\omega_1+2\omega_2
\end{alignedat}
\end{equation*}
and
\begin{equation*}\label{basesH2}
\begin{alignedat}{5}
H_2:&\quad
\omega_1=\tfrac{1}{5}((4+2\tau)\alpha_1+ (1+3\tau)\alpha_2),&\qquad\qquad
\alpha_1&=2\omega_1-\tau\omega_2,\,\ &\\
&\quad\omega_2=\tfrac15((1+3\tau)\alpha_1+ (4+2\tau)\alpha_2),&\qquad\qquad
\alpha_2&=-\tau\omega_1+2\omega_2.\\
\end{alignedat}
\end{equation*}

In all cases the  reflections $r_1$ and $r_2$  generate  $G$ are defined as follows
\begin{equation}\label{reflection}
r_k\lambda=\lambda-\frac{2\l \lambda\mid\alpha_k \r}{\l\alpha_k\mid\alpha_k\r}\alpha_k
\,,\qquad k=1,2,\quad \lambda\in\R^2.
\end{equation}
In particular,
\begin{equation*}
r_k0=0,\qquad
 r_k\omega_j=\omega_j-\delta_{jk}\alpha_k, \qquad
r_k\alpha_k=-\alpha_k.
\end{equation*}

An orbit of $G$ is the set of distinct points generated from a seed point $\lambda\in\R^2$ by repeated action of reflections \eqref{reflection}. Such an orbit contains at most as many points as is the order of $G$. %That is 6,\ 8,\ 12 and $10$ for $A_2$, $C_2$, $G_2$ and $H_2$ respectively.
Each orbit contains precisely one point with non-negative coordinates in $\omega$-basis.  The orbit $O(\lambda)$ is specified by this point, called the dominant point.

The orbits could be distinguished according to the position of their points, in particular of the dominant point. It is either point of the weight lattice $P$ or not in $\R^2$ (in crystallographic case).

The set of all dominants weights of $P$ is in non-negative  sector $P^+$, where
\begin{equation*}
  \begin{alignedat}{2}
P^+&=\Z^{\geq 0}\omega_1+\Z^{\geq 0}\omega_2\subset P&=\Z\omega_1+\Z\omega_2\subset\R^2.
\end{alignedat}
\end{equation*}
If $G$ stands for the non-crystallographic group $H_2$, the role of $\Z$ is played  by $\Z[\tau] $, quadratic extension of integer numbers by $\tau$, see \cite{P95,ChMP98}.

The size $|O(\lambda)|$ of an orbit $O(\lambda)$ is the number of distinct points it contains i.e.:
\begin{equation*}
  \begin{alignedat}{2}
|O(\lambda)|&=\left\{
\begin{array}{cll}
|G|             &\textrm{for}& \lambda=(a,b)\\
\tfrac12|G|     &\textrm{for}& \lambda=(a,0)\textrm{ or }\lambda=(0,b)\\
1               &\textrm{for}& \lambda=(0,0)
\end{array}
\right. ,\qquad
%|O(a,b)| &=|G|,  \\
%|O(a,0)|=|O(0,b)| &=\tfrac12|G|,\qquad a,b>0, \\
%|O(0,0)| &=1,
\end{alignedat}
\end{equation*}
where $ a,b>0.$

The product of two orbit functions of $G$ is the set of points obtained by adding to every point of one orbit  every point of the other orbit. Therefore the orbit sizes multiply,
$|O(\lambda)\otimes O(\lambda')|=|O(\lambda)|\cdot|O(\lambda')|$.
The products decompose into the sum of several orbits
\begin{equation}\label{decomposition}
O(\lambda)\otimes O(\lambda')=O(\lambda')\otimes O(\lambda)= O(\lambda+\lambda')\cup \cdots\cup mO(\lambda+\overline{\lambda'}),
\qquad m\in\mathbb N\,.
\end{equation}
Here $\overline{\lambda}$ stands for the lowest weight of the orbit $O(\lambda)$. If the sum $\lambda+\overline{\lambda'}$ is not a dominant weight, it should be reflected into the dominant weight of its orbit.% The multiplicity $m$ is strictly positive, $m\geq1$.

Except for the rank one of the group $G$, there has been no general formula for finding the terms in the decomposition \eqref{decomposition}. Such formulas are found in next paragraphs for the orbits of the four groups of rank 2.
\smallskip

Limits of orbits and product of orbits is another interesting subject which is briefly described below.

 Let one consider orbit $O(\lambda)$ of the group $G$ as a continues  function of two variables, in general from $\mathbb R^2$ to $\mathbb R^2$, where $\lambda=(x,y)=x\omega_1+y\omega_2$.  Taking into account that each orbit is characterized by dominant point  it is enough to calculate limits only for  points which coordinates are nonnegative real numbers. Limits  for this function  one could write in the form
\begin{equation}\label{limOrbit}
\begin{alignedat}{2}
\lim_{x\rightarrow a} O(x,y)&=\frac{l_1}{|O(a,y)|}O(a,y),\\
\lim_{y\rightarrow b} O(x,y)&=\frac{l_2}{|O(x,b)|}O(x,b),\\
\lim_{(x,y)\rightarrow (a,b)} O(x,y)&=\frac{6}{|O(a,b)|}O(a,b),
\end{alignedat}
\end{equation}
where  $a,b \geqslant 0$ and
$$
\begin{alignedat}{2}
l_1=\left\{
\begin{alignedat}{2}
6&\textrm{ for }y\neq 0\\
3&\textrm{ for }y=0
%1&\textrm{ for }x=y=0
\end{alignedat}\right. , \quad
&
l_2=\left\{
\begin{alignedat}{2}
6&\textrm{ for }x\neq 0\\
3&\textrm{ for }x=0
%1&\textrm{ for }x=y=0
\end{alignedat}\right. .
\end{alignedat}
$$
By analogy using \eqref{decomposition} and \eqref{limOrbit} one can calculate limits for product of two orbits
\begin{equation}\label{granice}
\begin{alignedat}{2}
  \lim_{x\rightarrow a} O(\lambda)\otimes O(\lambda')&=
  \lim_{x\rightarrow a} O(\lambda+\lambda')\cup \cdots\cup  \lim_{x\rightarrow a} mO(\lambda+\overline{\lambda'}),\\
  \lim_{y\rightarrow b} O(\lambda)\otimes O(\lambda')&=
    \lim_{y\rightarrow b} O(\lambda+\lambda')\cup \cdots\cup  \lim_{y\rightarrow b} mO(\lambda+\overline{\lambda'}),\\
  \lim_{(x,y)\rightarrow (a,b)}O(\lambda)\otimes O(\lambda')&=\!\!\!\!\lim_{\!\!\!\! (x,y)\rightarrow (a,b)}\!\! O(\lambda+\lambda')\cup \cdots\cup  \lim_{(x,y)\rightarrow (a,b)} mO(\lambda+\overline{\lambda'}),\\
\end{alignedat}
\end{equation}
where multiplicity of orbits are also functions of $x$ and $y.$ When $a$ or $b$ are equal $0$ then the limits are right side limits.

Some examples of limits are presented  in appropriate paragraphs.

%%%%%%%%%%%%%%%%%%%%%%%%%%%%%%%%%%%%%%%%%%%%%%%%%%%%%%%%%%%%%
\section{Decompositions of products of orbits of $A_2$}\label{A2}

There is a useful general symmetry property of all orbits $O(\lambda)$ of $A_2$, where $\lambda=(a,b)=a\omega_1+b\omega_2$:
\begin{gather}\label{outer}
O(a,b) = -O(b,a)\,,\qquad a,b\in\R\,.
\end{gather}
In most cases we are interested in $a,b\in\Z$, however \eqref{outer} is valid for any real $a$ and $b$. For every point  $\lambda\in O(a,b)$ there is the point $-\lambda\in O(b,a)$. The property is a consequence of the outer automorphism of $A_2$.

Another useful general hierarchy of orbits $O(a,b)$ of $A_2$ with integers $a$ and $b$, is their splitting into three mutually exclusive congruence classes according to the value of their congruence number $K(a,b)$
\begin{gather*}
K(a,b)\equiv 2a+b\pmod3\,,\qquad a,b\in\Z\,.
\end{gather*}
All points of an orbit are in the same congruence class. It is the consequence of the fact that difference between two points of the same orbit is an integer linear combination of simple roots. All simple roots belong to the same congruence class, namely 0. During the multiplication of orbits, their congruence numbers  add up. All orbits in the decomposition belong to that congruence class, see \cite{KP06, LP82}.

There are four types of orbits $O(a,b)$ for this group, see for example \cite{KP06}:
\begin{equation*}
\begin{alignedat}{2}
O(0,0)&=\{(0,0)\},\quad \\
O(a,0)&=\{(a,0),(-a,a),(0,-a)\},\qquad\quad
O(0,b)=\{(0,b),(b,-b),(-b,0)\}, \\
O(a,b)&=\{(a,b),(-a, a+b),(a+b,-b),(-a-b,a),(b,-a-b),(-b,-a)\}.\\
\end{alignedat}
\end{equation*}

For $\lambda=(a_1,a_2)=a_1\omega_1+a_2\omega_2$ and $\lambda'=(b_1,b_2)=b_1\omega_1+b_2\omega_2$ using duality of bases, see \eqref{weightreflect}, one can get following relations:
\begin{equation}\label{zmienneA2}
\begin{alignedat}{2}
\langle \lambda\mid \alpha_j\rangle
& =a_j, \\
\langle\lambda'\mid \alpha_j\rangle
& =b_j,\\
\end{alignedat}\qquad\textrm{ for }j=1,2\textrm{ and }a_1,a_2,b_1,b_2\in\R^{\geq0}
\end{equation}
 which are  used in the proposition below.

\medskip
\begin{prop}\label{A2decomposition}\

Decomposition of the product \eqref{decomposition} of two orbits of $A_2$  with dominant weights $\lambda=(a_1,a_2)$ and $\lambda'=(b_1,b_2)$  for $a_1,a_2,b_1,b_2\in\R^{\geq0}$ is given by the following formula
%%%%%%%%%%%%%%%%%%%%%%%%%%%%%%%%%%%%%

\smallskip
\begin{eqnarray}\label{A2product}
&&\!\!\!\!\!\!\!\!\!\!\!\!O(\lambda)\otimes O(\lambda')
 \\
\nonumber&&\!\!\!\!\!\!\!\!\!\!\!\!\!\qquad = k_1\;O(\lambda+\lambda')
%\qquad\qquad\qquad\qquad\qquad\qquad a_1,a_2,b_1,b_2\in\Z^{\geq0},
\\
\nonumber&&\!\!\!\!\!\!\!\!\!\!\!\!\qquad\cup k_2\;O\left(\Big|\langle\lambda-\lambda'\mid \a_1\rangle \Big|,\;\langle \lambda+\lambda'\mid \tfrac12\a_1+\a_2\rangle-\tfrac12\Big|\langle\lambda-\lambda'\mid\a_1\rangle\Big| \right)
\\
\nonumber&&\!\!\!\!\!\!\!\!\!\!\!\!\qquad\cup k_3\; O\left( \langle\lambda+\lambda'\mid\a_1+\tfrac12\a_2\rangle-\tfrac12\Big|\langle\lambda-\lambda'\mid\a_2\rangle\Big| , \; \Big|\langle\lambda-\lambda'\mid\a_2\rangle \Big|\right)
\\
\nonumber&&\!\!\!\!\!\!\!\!\!\!\!\!\qquad\cup k_4\;O\left(\Big|\langle \lambda'\mid\a_1+\tfrac12\a_2\rangle+\tfrac12\langle\lambda\mid\a_2-\a_1\rangle-\tfrac12\big|\langle\lambda-\lambda'\mid\a_2\rangle
+\langle\lambda\mid\a_1 \rangle\big|\Big|, \right.
\\
\nonumber&&\!\!\!\!\!\!\!\!\!\!\!\!\!\!\qquad\qquad\qquad\left.\Big|\langle\lambda\mid\tfrac12\a_1+\a_2 \rangle+\tfrac12\langle\lambda'\mid\a_1-\a_2\rangle -\tfrac12\big|\langle\lambda'-\lambda\mid\a_1\rangle+\langle\lambda'\mid\a_2\rangle\big|\Big|\right)
\\
\nonumber&&\!\!\!\!\!\!\!\!\!\!\!\! \qquad\cup k_5\;O\left(\Big|\langle \lambda\mid\a_1+\tfrac12\a_2\rangle+\tfrac12\langle\lambda'\mid\a_2-\a_1\rangle-\tfrac12\big|\langle\lambda'-\lambda\mid\a_2\rangle
+\langle\lambda'\mid\a_1\rangle\big|\Big|, \right.
\\
\nonumber&&\!\!\!\!\!\!\!\!\!\!\!\!\!\!\qquad\qquad\qquad\left.\Big|\langle\lambda'\mid\tfrac12\a_1+\a_2\rangle+
\tfrac12\langle\lambda\mid\a_1-\a_2\rangle -\tfrac12\big|\langle\lambda-\lambda'\mid\a_1\rangle+\langle\lambda\mid\a_2\rangle\big|\Big|\right)
\\
\nonumber&&\!\!\!\!\!\!\!\!\!\!\!\cup k_6 \;O\left(\Big|\big|\langle \lambda\!-\!\lambda'\mid\a_1+\a_2\rangle \big|\!-\!\big|\!\min\{\langle \lambda\mid\a_1\rangle-\langle\lambda'\mid\a_2\rangle,\langle\lambda'\mid\a_1\rangle-\langle\lambda\mid\a_2\rangle,0\}\big|\Big|\right.,
\\
\nonumber&&\!\!\!\!\!\!\!\!\!\!\!\!\!\!\quad\left.\Big|\big|\langle\lambda-\lambda'\mid \a_1+\a_2\rangle \big|\!-\!\big|\!\min\{-\langle \lambda\mid\a_1\rangle+\langle\lambda'\mid\a_2\rangle,-\langle\lambda'\mid\a_1\rangle+\langle\lambda\mid\a_2\rangle,0\}
\big|\Big|\right)
\end{eqnarray}
where the multiplicities $k_1,\dots,k_6$ are given in terms of orbits sizes by:
{\footnotesize \begin{align*}
&k_1= \tfrac16\tfrac{|O(\lambda)||O(\lambda')|}{|O( \lambda+\lambda')|}
\\
&k_2= \tfrac16\tfrac{|O(\lambda)||O(\lambda')|}
     {\Big|O\left(\big|\langle\lambda-\lambda'\mid \a_1\rangle \big|,\;\langle \lambda+\lambda'\mid \tfrac12\a_1+\a_2\rangle-\tfrac12\big|\langle\lambda-\lambda'\mid\a_1\rangle\big| \right) \Big|}
\\
&k_3=\tfrac16\tfrac{|O(\lambda)||O(\lambda')|}{\Big|O\left( \langle\lambda+\lambda'\mid\a_1+\tfrac12\a_2\rangle-\tfrac12\big|
\langle\lambda-\lambda'\mid\a_2\rangle\big| , \; \big|\langle\lambda-\lambda'\mid\a_2\rangle \big|\right)\Big|}\\
&k_4= \tfrac16
|O(\lambda)|
  |O(\lambda')|\Big/\Big|O\left(\Big|\langle \lambda'\mid\a_1+\tfrac12\a_2\rangle+\tfrac12\langle\lambda\mid\a_2-\a_1\rangle-\tfrac12\big|\langle\lambda-\lambda'\mid\a_2\rangle
+\langle\lambda\mid\a_1 \rangle\big|\Big|, \right.
\\
&\qquad\qquad\qquad\qquad \left.\Big|\langle\lambda\mid\tfrac12\a_1+\a_2 \rangle+\tfrac12\langle\lambda'\mid\a_1-\a_2\rangle -\tfrac12\big|\langle\lambda'-\lambda\mid\a_1\rangle+\langle\lambda'\mid\a_2\rangle\big|\Big|\right)\Big|\\
&k_5= \tfrac16 |O(\lambda)||O(\lambda')|\Big/\Big| O\left(\Big|\langle \lambda\mid\a_1+\tfrac12\a_2\rangle+\tfrac12\langle\lambda'\mid\a_2-\a_1\rangle-\tfrac12\big|\langle\lambda'-\lambda\mid\a_2\rangle
+\langle\lambda'\mid\a_1\rangle\big|\Big|, \right.
\\
&\qquad\qquad\qquad\qquad \left.\Big|\langle\lambda'\mid\tfrac12\a_1+\a_2\rangle+
\tfrac12\langle\lambda\mid\a_1-\a_2\rangle -\tfrac12\big|\langle\lambda-\lambda'\mid\a_1\rangle+\langle\lambda\mid\a_2\rangle\big|\Big|\right) \Big|\\
&k_6=\tfrac16{|O(\lambda)||O(\lambda')|}\cdot\\
&\qquad\qquad 1\Big/\Big| O(\Big|\big|\langle \lambda-\lambda'\mid\a_1+\a_2\rangle \big|-\big|\min\{\langle \lambda\mid\a_1\rangle-\langle\lambda'\mid\a_2\rangle,\langle\lambda'\mid\a_1\rangle-\langle\lambda\mid\a_2\rangle,0\}\big|\Big|
\\
&\qquad\qquad\qquad \Big|\big|\langle\lambda-\lambda'\mid \a_1+\a_2\rangle \big|-\big|\min\{-\langle \lambda\mid\a_1\rangle+\langle\lambda'\mid\a_2\rangle,-\langle\lambda'\mid\a_1\rangle+\langle\lambda\mid\a_2\rangle,0\}
\big|\Big|) \Big|.
\end{align*}
}

\end{prop}

Sketch of the prove is given in the Appendix A.

Note that the congruences class of each  term in the decomposition \eqref{A2product} coincide. Namely, the congruence class, which is the sum of congruence classes of factors of $O(a_1,a_2)\otimes O(b_1,b_2).$
Lets illustrate this fact in an example.

\begin{example}\
Using \eqref{A2product} one has:
\begin{equation}\label{A2example1}
\begin{alignedat}{2}
    O(1, 2)\otimes O(3,1) &=O(4, 3)\cup O(2, 4) \\
    &\qquad\cup O(3, 2) \cup O(5, 1)\cup 2 O(0, 2)\cup 2O(1, 0).
  \end{alignedat}\end{equation}
Then it is easy to check that
\begin{align*}
  &K(1,2)+K(3,1)%=1\pmod3+1\pmod3
=2\pmod3,\\
  &K(4, 3)= K(2, 4)=K(3, 2)=  K(5, 1)=K(0, 2)=  K(1,0)=2\pmod3.
\end{align*}
Orbits and decomposition of  product \eqref{A2example1} are presented in figure \ref{RysA2}.
\begin{figure}[h]\label{RysA2}
\begin{center}
\subfigure[][]{\label{RysA2a}\includegraphics[scale=0.35]{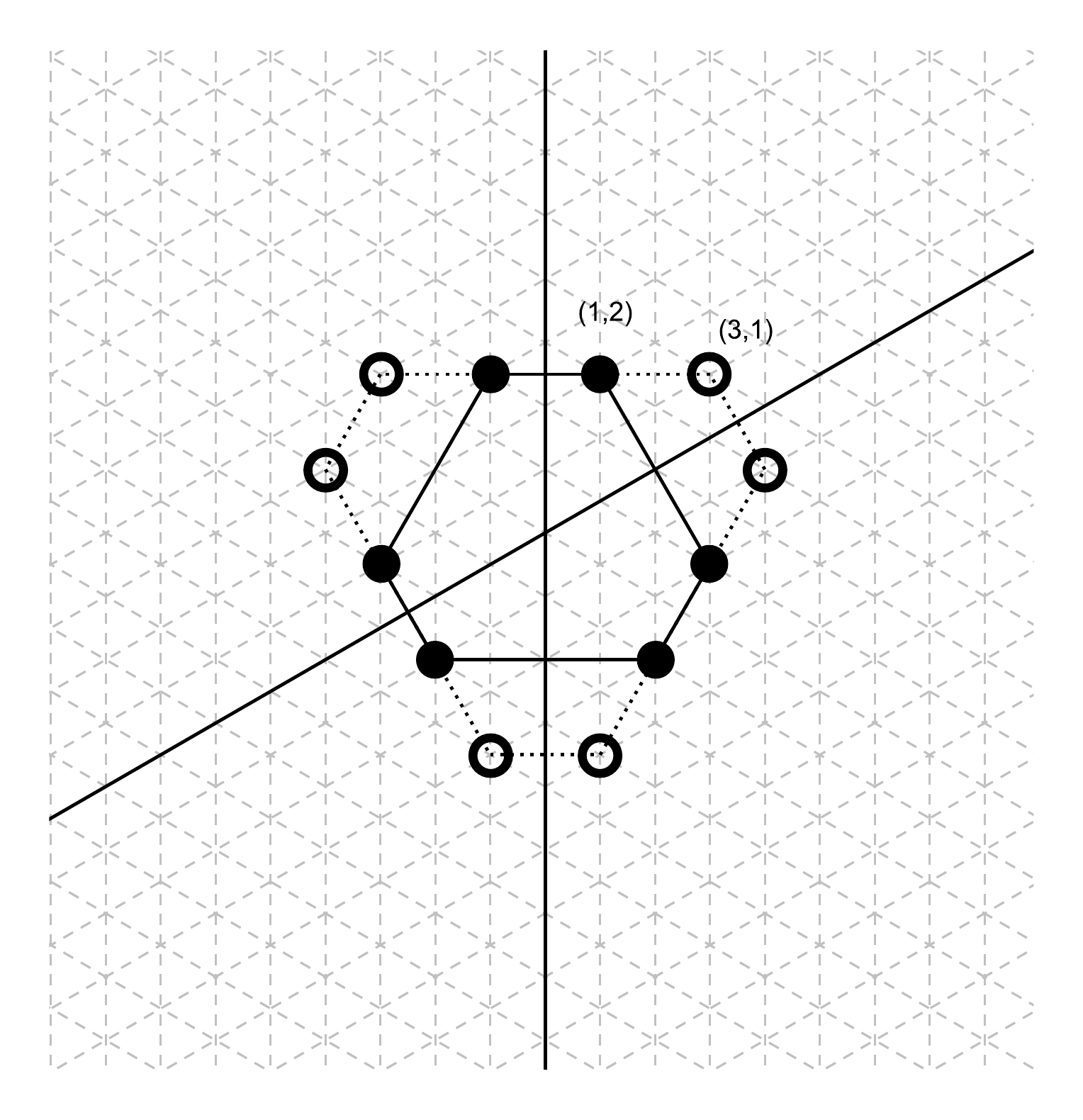}}
\;
\subfigure[][]{\label{RysA2b}\includegraphics[scale=0.35]{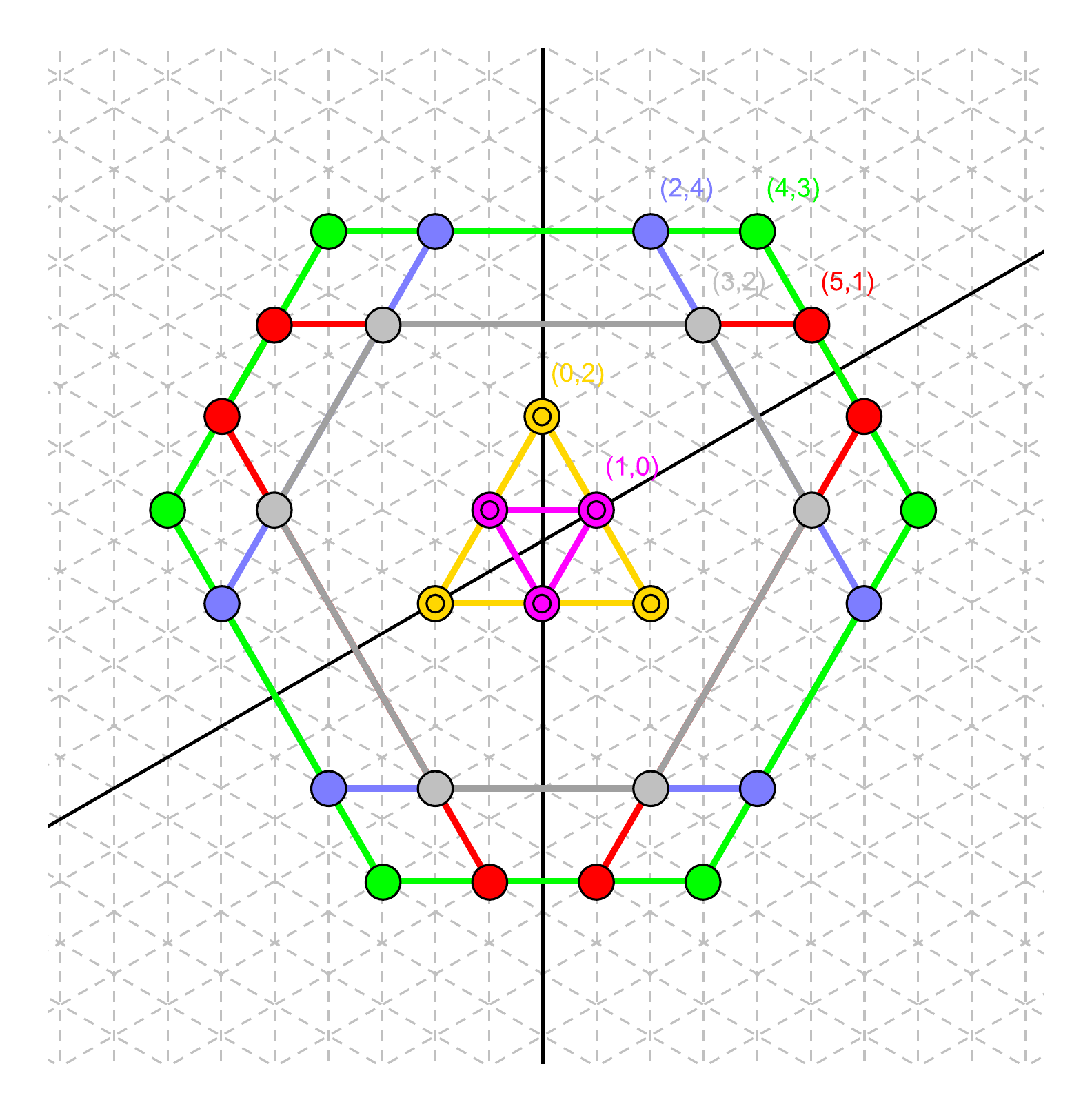}}
\end{center}
\caption{\subref{RysA2a} the six points of the orbit $O(1,2)$ and of $O(3,1)$ of $A_2$ are presented. The straight lines are the reflection mirrors containing $\omega_1$ and $\omega_2$. The dominant points are indicated in the positive sector. Dashed lines are the directions of the weight lattice axes of $A_2$.\newline
 \subref{RysA2b} The six orbits of the decomposition of the product $O(1,2)\otimes O(3,1)$ of $A_2$. Four of them are hexagons and two are triangles taken twice what is denoted by double dots. The straight lines are the reflection mirrors containing $\omega_1$ and $\omega_2$. The dominant points are indicated in the positive sector. Dashed lines are the directions of the weight lattice axes of $A_2$. }
\end{figure}

Using \eqref{granice} one can get
  \begin{equation*}
 \lim_{y\rightarrow 0^+} O(1,2)\otimes O(3,y)=2O(4, 2)\cup 2O(2, 3) \cup 4O(0,1)=O(1,2)\otimes 2O( 3,0).
  \end{equation*}
In  figure  \ref{limitEX1}  graphical interpretation of this example is shown. Others possible limits could be calculated and drown in the same way.
\begin{figure}[h]
\begin{center}
\subfigure[][]{\label{limitEX1a}
\includegraphics[scale=0.35]{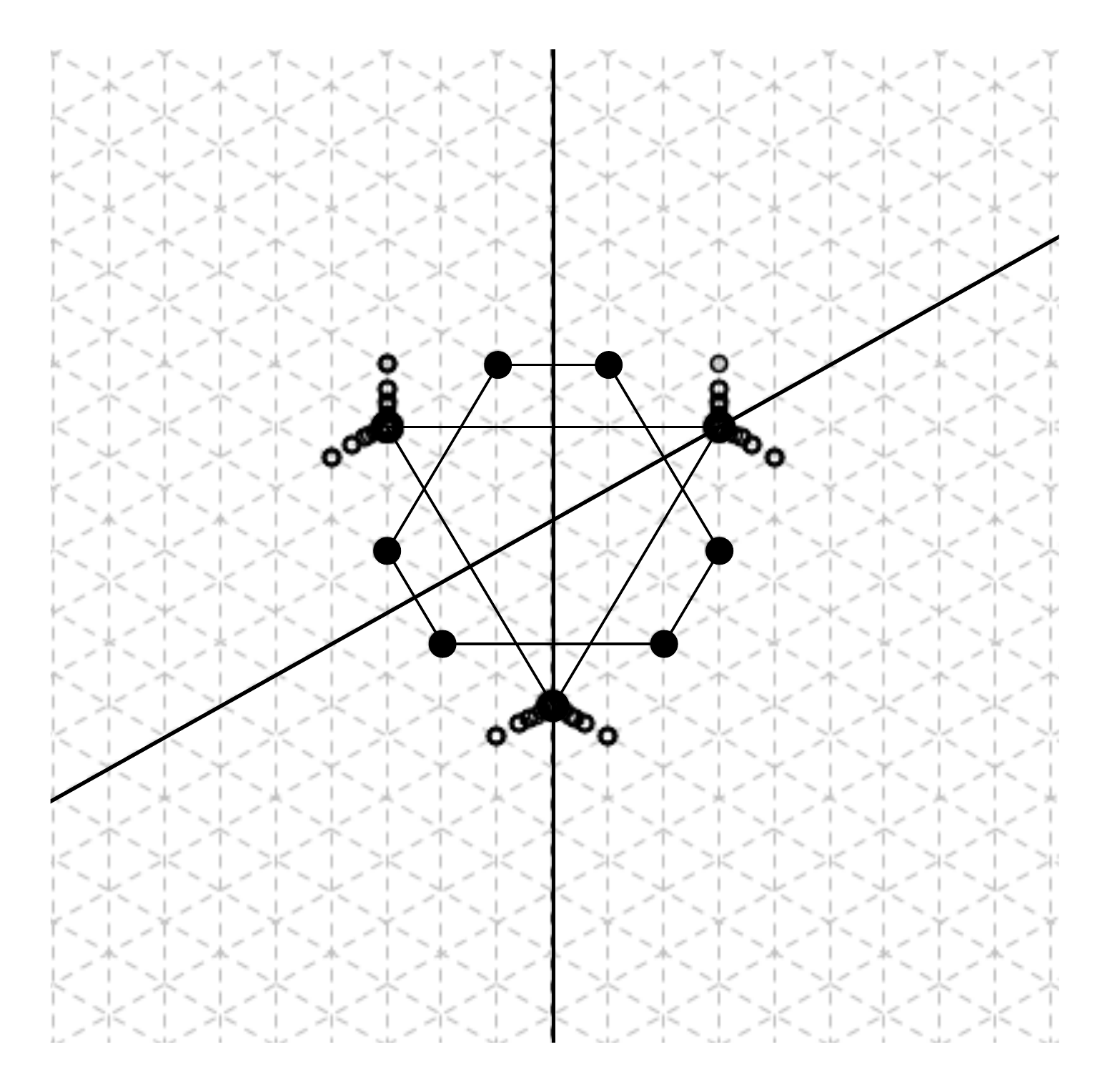}}
\;
\subfigure[][]{\label{limitEX1b}
\includegraphics[scale=0.35]{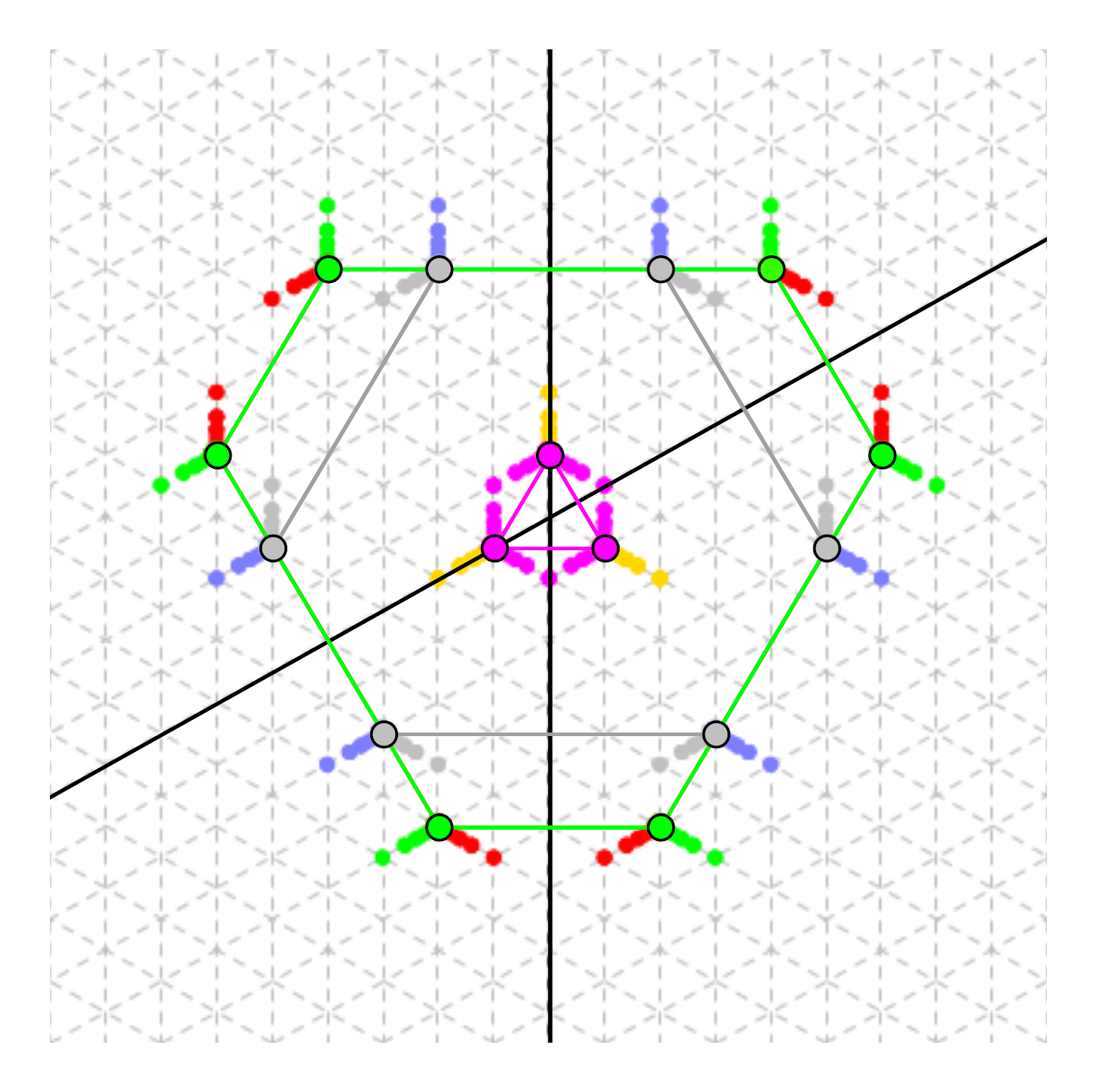}}\end{center}
\caption{\subref{limitEX1a} Orbits $ O(1, 2)\;$ and $\lim\limits_{y\rightarrow 0^+} O(3,y) $.  \newline
\subref{limitEX1b} Decomposition of product of $ \lim\limits_{y\rightarrow 0^+}O(1, 2)\otimes O(3,y)$.}\label{limitEX1}
\end{figure}

\end{example}
%%%%%%%%%%%%%%%%%%%%%%%%%%%%%%%%%%%%%%%%%%%%%%%%%%%%%%%%%%%%%
\section{Decompositions of products of orbits of  $C_2$}\label{C2}

For the group $C_2$ is also a general symmetry property of all orbits of this group:
\begin{gather}\label{autoC}
\lambda\in O(a,b)\quad \Longleftrightarrow\quad -\lambda\in O(a,b)\,,\qquad a,b\in\R\,.
\end{gather}
In most cases we are interested in $a,b\in\Z$, however \eqref{autoC} is valid for any real $a$ and $b$.

A useful general hierarchy of orbits $O(a,b)$ of $C_2$ with integer $a$ and $b$, is their splitting into two mutually exclusive congruence classes according to the value of their congruence number $K(a,b)$
\begin{gather*}
K(a,b)=a\pmod 2\,,\qquad a,b\in\Z\,.
\end{gather*}
All points of an orbit are in the same congruence class. It is the consequence of the fact that difference between two points of the same orbit is an integer linear combination of simple root, and all simple roots are in the congruence class 0. During the multiplication of orbits, their congruence numbers  add up. All orbits in the decomposition belong to that congruence class.

\smallskip
There are also four kinds of   orbits for this group,one can find it, for example in \cite{KP06,LP82}:
\begin{alignat*}{2}
O(0,0)&=\{(0,0)\},\\
O(a, 0)&=\{\pm(a, 0), \pm(-a, a)\},\\
O(0, b)&=\{ \pm(0 ,b), \pm(2b, -b)\},\\
O(a, b)&=\{ \pm(a, b), \pm(-a, a+b), \pm(a+2b, -b), \pm(a+2b, -a-b)\}.
\end{alignat*}

Let $\lambda=(a_1,a_2)=a_1\omega_1+a_2\omega_2$ and $\lambda'=(b_1,b_2)=b_1\omega_1+b_2\omega_2$. Using duality of bases one can get following relations:
\begin{equation}\label{zmienneC2}
\begin{alignedat}{2}
\langle\lambda\mid\alpha_1\rangle&=\tfrac12a_1, \qquad& \langle\lambda'\mid\alpha_1\rangle &=\tfrac12b_1,\\
\langle\lambda\mid\alpha_2\rangle &=a_2, \qquad&\langle\lambda'\mid\alpha_2\rangle &=b_2, \\
\end{alignedat}\qquad\textrm{ for }a_1,a_2,b_1,b_2\in\R^{\geq0}
\end{equation}
 which are  useful in the proposition below.
\begin{prop}\label{C2decomposition}\

Decomposition of the product \eqref{decomposition} of two  orbits of $C_2$  with dominant weights $\lambda=(a_1,a_2)$ and $\lambda'=(b_1,b_2)$ for $a_1,a_2,b_1,b_2\in\R^{\geq0}$ is given by the following formula
%%%%%%%%%%%%%%%%%%%%%%%%%%%%%%%%%%%%%
\begin{equation}
\begin{aligned}\label{C2product}
&O(\lambda)\otimes O(\lambda')
 \\
&\qquad = k_1\;O(\lambda+\lambda')%\qquad\qquad\qquad\qquad\qquad\qquad a_1,a_2,b_1,b_2\in\Z^{\geq0}
\\
&\qquad\cup k_2\;O(|\langle \lambda-\lambda'\mid2\a_1 \rangle|,\;\langle \lambda+\lambda'\mid\a_1+\a_2\rangle-|\langle \lambda-\lambda'\mid\a_1 \rangle|)
\\
&\qquad\cup
  k_3\;O(\langle\lambda+\lambda'\mid 2\a_1+\a_2\rangle-|\langle \lambda-\lambda'\mid\a_2\rangle|,\;|\langle \lambda-\lambda'\mid\a_2\rangle|)
\\
& \qquad\cup
 k_4\;O\left(|\langle\lambda+\lambda'\mid\a_2\rangle+\langle\lambda\mid2\a_1\rangle- |\langle \lambda'\mid 2\a_1\rangle+\langle\lambda'-\lambda\mid\a_2\rangle||,\right.\\
 &\qquad\qquad\qquad\qquad\left.|\langle\lambda+\lambda'\mid\a_1\rangle+\langle\lambda' \mid\a_2 \rangle-|\langle\lambda'-\lambda\mid\a_1\rangle-\langle\lambda\mid\a_2\rangle||\right)
 \\
& \qquad\cup
k_5\;O\left(\min\{\langle2\lambda\mid\a_1+\a_2\rangle+\langle\lambda'\mid 2\a_1\rangle, |\langle\lambda\mid2\a_1\rangle\!-\!\langle2\lambda'\mid\a_1+\a_2\rangle|\},\right.\\
&\qquad\qquad\qquad\qquad\left.|\langle\lambda+\lambda'\mid\a_1\rangle+\langle\lambda\mid\a_2\rangle -|\langle\lambda-\lambda'\mid\a_1\rangle-\langle\lambda'\mid\a_2\rangle||\right)
\\
&\qquad\cup k_6\;O\left(2\min\{|\langle\lambda-\lambda'\mid \a_1\rangle|, |\langle\lambda-\lambda'\mid\a_1+\a_2\rangle|\},\right.\\
&\qquad\qquad\qquad\qquad\qquad\left.\min\{|\langle\lambda-\lambda'\mid\a_2\rangle|, |\langle\lambda-\lambda'\mid2\a_1+\a_2\rangle|\}\right)
\\
&\qquad\cup
k_7\;O\left(2\min\{\langle\lambda+\lambda'\mid\a_1\rangle,|\langle\lambda-\lambda'\mid\a_1+\a_2\rangle|\}, \right.\\
&\qquad\qquad\qquad\qquad\qquad\left.|\langle\lambda+\lambda'\mid\a_1\rangle-|\langle\lambda'-\lambda\mid\a_1+ \a_2\rangle||\right)
\\
&\qquad\cup
 k_8\;O\left(|\langle\lambda+\lambda'\mid\a_2\rangle-|\langle\lambda'-\lambda\mid2\a_1+\a_2\rangle||,\right.\\
&\qquad\qquad\qquad\qquad\qquad\left.
\min\{\langle\lambda+\lambda'\mid\a_2\rangle,|\langle\lambda-\lambda'\mid2\a_1+\a_2\rangle|\}\right)
\end{aligned}
\end{equation}
where the multiplicities $k_1,\dots,k_8$ are given in terms of orbits sizes by:
{\footnotesize
\begin{align*}
&k_1=\tfrac18\frac{|O(\lambda)||O(\lambda')|}{|O(\lambda+\lambda')|}\\
&k_2=\tfrac18\frac{|O(\lambda)||O(\lambda')|}{|O(|\langle \lambda-\lambda'\mid2\a_1 \rangle|,\;\langle \lambda+\lambda'\mid\a_1+\a_2\rangle-|\langle \lambda-\lambda'\mid\a_1 \rangle|)|}\\
&k_3=\tfrac18\frac{|O(\lambda)||O(\lambda')|}{|O(\langle\lambda+\lambda'\mid 2\a_1+\a_2\rangle-|\langle \lambda-\lambda'\mid\a_2\rangle|,\;|\langle \lambda-\lambda'\mid\a_2\rangle|)|}\\
&k_4=\tfrac18 |O(\lambda)||O(\lambda')|\Big/\Big|O\left(|\langle\lambda+\lambda'\mid\a_2\rangle+\langle\lambda\mid2\a_1\rangle- |\langle \lambda'\mid 2\a_1\rangle+\langle\lambda'-\lambda\mid\a_2\rangle||,\right.
\\
&\qquad\qquad \left.|\langle\lambda+\lambda'\mid\a_1\rangle+\langle\lambda' \mid\a_2 \rangle-|\langle\lambda'-\lambda\mid\a_1\rangle-\langle\lambda\mid\a_2\rangle||\right)\Big|\\
&k_5=\tfrac18|O(\lambda)||O(\lambda')|\Big/\Big|O\left(\min\{\langle2\lambda\mid\a_1+\a_2\rangle+\langle\lambda'\mid 2\a_1\rangle, |\langle\lambda\mid2\a_1\rangle-\langle2\lambda'\mid\a_1+\a_2\rangle|\},\right.
\\
&\qquad\qquad \left.|\langle\lambda+\lambda'\mid\a_1\rangle+\langle\lambda\mid\a_2\rangle -|\langle\lambda-\lambda'\mid\a_1\rangle-\langle\lambda'\mid\a_2\rangle||\right) \Big|\\
&k_6=\tfrac18 |O(\lambda)||O(\lambda')|
\\
& 1\Big/|O\left(2\min\{|\langle\lambda-\lambda'\mid \a_1\rangle|, |\langle\lambda-\lambda'\mid\a_1+\a_2\rangle|\},\min\{|\langle\lambda-\lambda'\mid\a_2\rangle|, |\langle\lambda-\lambda'\mid2\a_1+\a_2\rangle|\}\right)\!|\\
&k_7=\tfrac18\frac{|O(\lambda)||O(\lambda')|}{|O\left(2\min\{\langle\lambda+\lambda'\mid\a_1\rangle,|\langle\lambda-\lambda'\mid\a_1+\a_2\rangle|\}, \;|\langle\lambda+\lambda'\mid\a_1\rangle-|\langle\lambda'-\lambda\mid\a_1+ \a_2\rangle||\right)|}\\
&k_8=\tfrac18\frac{|O(\lambda)||O(\lambda')|}{|O\left(|\langle\lambda+\lambda'\mid\a_2\rangle-|\langle\lambda'-\lambda\mid2\a_1+\a_2\rangle||,\;
\min\{\langle\lambda+\lambda'\mid\a_2\rangle,|\langle\lambda-\lambda'\mid2\a_1+\a_2\rangle|\}\right)|}\;.
\end{align*}}
\end{prop}

\medskip

The formula \eqref{C2product} can be verified in the same way as \eqref{A2product}.

Analogously  sum of congruence classes of the product factors is equal the congruence class of each  term in the decomposition \eqref{C2product}, which is easy to verify by straight forward calculation of congruence classes.
Let illustrate this fact in an example.

\begin{example}\
Using \eqref{zmienneC2}and proposition \ref{C2decomposition} for $\lambda=(1,2)$ and $\lambda'=(3,1)$ one gets
\begin{equation}\label{C2example}
\begin{alignedat}{1}
O(1,2)\otimes O(3,1)&=O(4, 3) \cup O(6, 1)\cup  O(4,2)\\
&\qquad  \cup O(2,4)\cup O(2,2)\cup O(2,1)\cup 2O(0,2)\cup 2O(0,1)
\end{alignedat}
\end{equation}
and
\begin{equation*}
  \begin{alignedat}{1}
K(1,2)+K(3,1)&=0\pmod 2=K(4,3)=K(6,1)=K(4,2)\\
&=K(2,4)=K(2,2)=K(2,1)=K(0,2)=K(0,1).
\end{alignedat}
\end{equation*}
Orbits and decomposition of  product \eqref{C2example} are presented in figure \ref{rysC2}.  More advanced examples one could find in appendix  B.
\begin{figure}[h]
\begin{center}
\subfigure[][]{\label{rysC2a}\includegraphics[scale=0.47]{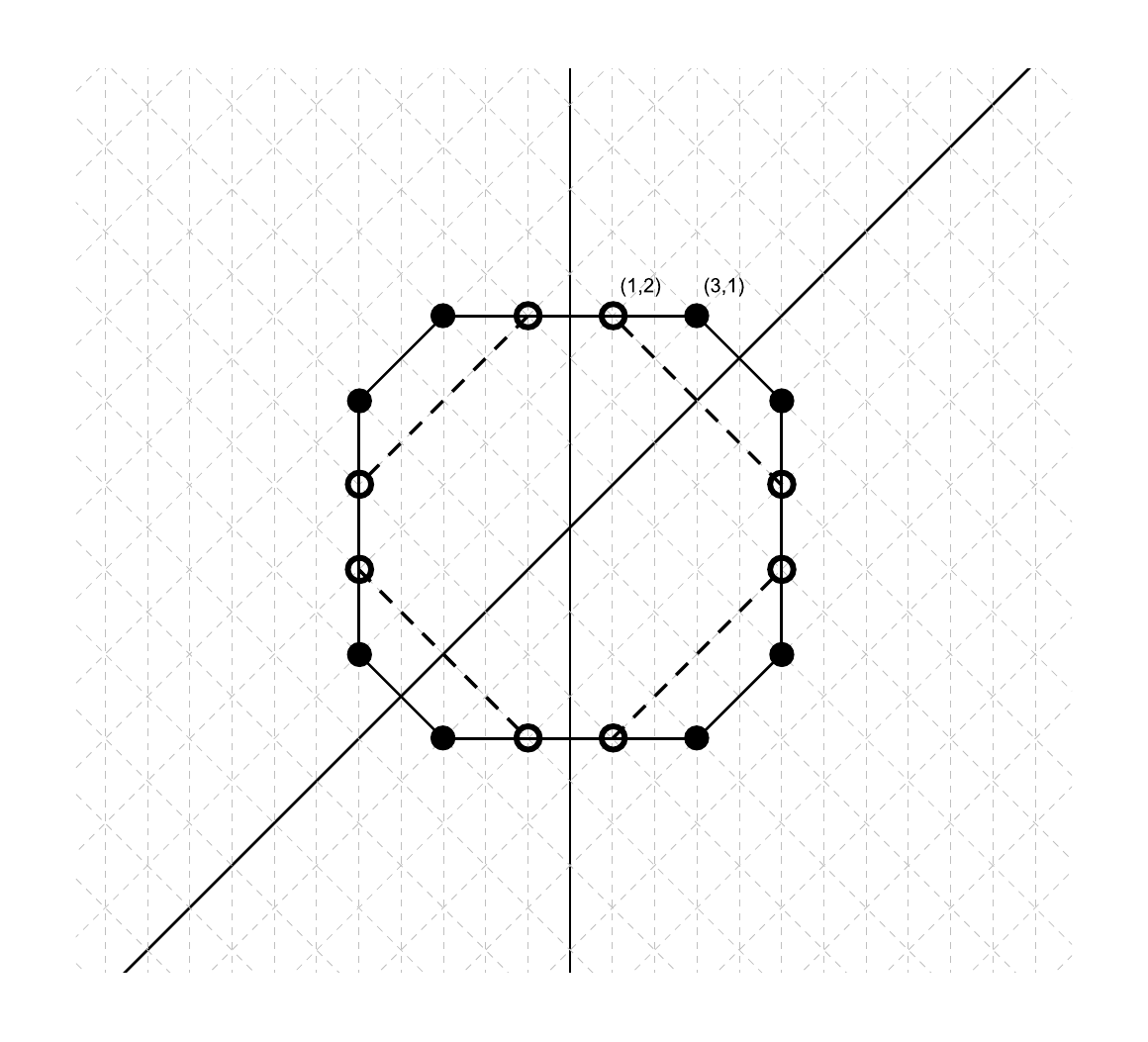}}
 \;
\subfigure[][]{\label{rysC2b}
\includegraphics[scale=0.47]{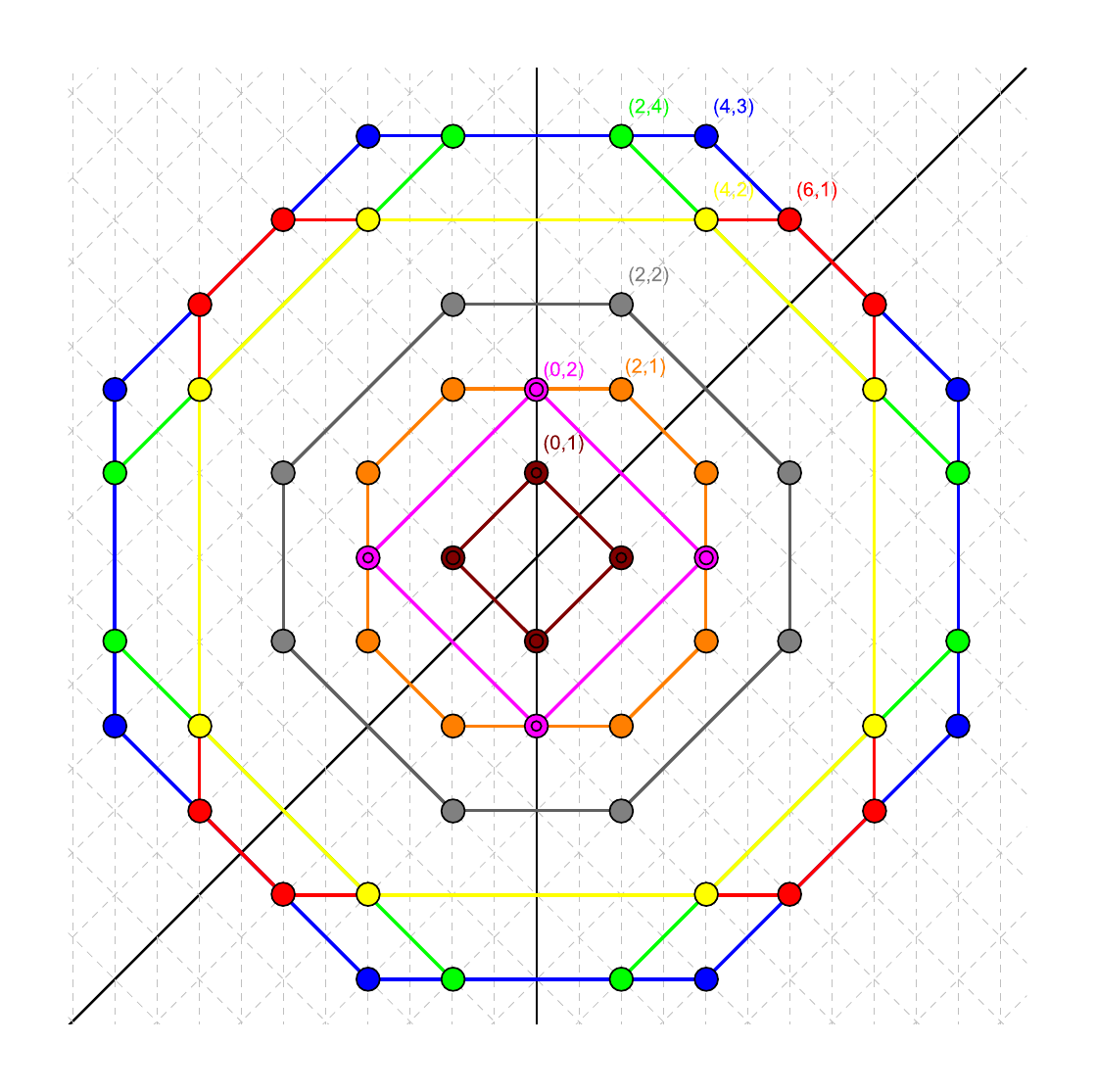}}
\caption{\subref{rysC2a} The eight points of the orbit $O(1,2)$ and of $O(3,1)$ of $C_2$. The straight lines are the reflection mirrors containing $\omega_1$ and $\omega_2$. The dominant points are indicated in the positive sector. Dashed lines are the directions of the weight lattice axes of $C_2$. \newline
\subref{rysC2b} The eight orbits of the decomposition of the product $O(1,2)\otimes O(3,1)$ of $C_2$. Six of them are octagons and two are squares taken twice what is denoted by double dots. The straight lines are the reflection mirrors containing $\omega_1$ and $\omega_2$. The dominant points are indicated in the positive sector. Dashed lines are the directions of the weight lattice axes of $C_2$.}\label{rysC2}
\end{center}
\end{figure}\

Using \eqref{granice} one can get
\begin{equation*}
\begin{alignedat}{1}
& \lim_{x\rightarrow 2}O(1,2)\otimes O(x,1)\\
&\;  =
O(3, 3)\cup O(5, 1)\cup O(3, 2)\cup O(1, 4) \cup O(3, 1)\cup2 O(3, 0)\cup O(1, 1)\cup 2 O(1, 0)\\
&\;  =O(1,2)\otimes O(2,1).
\end{alignedat}
\end{equation*}

\end{example}
%%%%%%%%%%%%%%%%%%%%%%%%%%%%%%%%%%%%%%%%%%%
\section{Decompositions of products of orbits of $G_2$}

There is also a general symmetry property of all orbits of $G_2$:
\begin{gather}\label{autoG}
\lambda\in O(a,b)\quad \Longleftrightarrow\quad -\lambda\in O(a,b)\,,\qquad a,b\in\R\,.
\end{gather}
In most cases we are interested in $a,b\in\Z$, however \eqref{autoG} is valid for any real $a$ and~$b$.  All weights are in the same congruence class.

For this group one has also three kinds of  nontrivial orbits and one trivial, see for example \cite{KP06,LP82}:
\begin{alignat*}{2}
O(0,0)&=\{(0,0)\},\\
O(a,0)&=\{ \pm(a ,0), \pm(-a ,3a), \pm(2a ,-3a)\},\\
O(0,b)&=\{ \pm(0 ,b), \pm(b, -b), \pm(-b, 2b), \},\\
O(a,b)&=\left\{ \pm(a, b), \pm(-a, 3a+b), \pm(a+b, -b),\right.\\
      &\left.\pm (2a+b, -3a-b),\pm(-a-b, 3a+2b), \pm(-2a-b, 3a+2b)\right\}.
\end{alignat*}
As before product of two orbits of $G_2$ is written in terms of \break$\lambda=(a_1,a_2)=a_1\omega_1+a_2\omega_2$ and $\lambda'=(b_1,b_2)=b_1\omega_1+b_2\omega_2$ and simple roots, i.e.:
\begin{equation*}
\begin{alignedat}{2}
\langle\lambda\mid\alpha_1\rangle&=a_1, \qquad& \langle\lambda'\mid\alpha_1\rangle &=b_1,\\
\langle\lambda\mid\alpha_2\rangle &=\tfrac13a_2, \qquad&\langle\lambda'\mid\alpha_2\rangle &=\tfrac13b_2, \\
\end{alignedat}\textrm{ for }a_1,a_2,b_1,b_2\in\R^{\geq0}.
\end{equation*}

\begin{prop}\label{G2decomposition}\

Decomposition of the product \eqref{decomposition} of two   orbits of $G_2$  with dominant weights $\lambda=(a_1,a_2)$ and $\lambda'=(b_1,b_2)$ for $a_1,a_2,b_1,b_2\in\R^{\geq0}$ is given by the following formula
%%%%%%%%%%%%%%%%%%%%%%%%%%%%%%%%%%%%%
{
\begin{eqnarray}\label{G2product}
&&O(\lambda)\otimes O(\lambda')
 \\
\nonumber&&\quad = k_1 O(\lambda+\lambda') %\qquad\qquad\qquad\qquad\qquad\qquad a_1,a_2,b_1,b_2\in\Z^{\geq0}
\\
\nonumber&&\quad\cup   k_2 O(|\langle\lambda-\lambda'\mid\a_1\rangle|,\tfrac32(\langle\lambda+\lambda'\mid\a_1+2\a_2\rangle- |\langle\lambda-\lambda'\mid\a_1\r|))
\\
\nonumber&&\quad\cup k_3 O(\langle\lambda+\lambda'\mid \a_1+\tfrac32\a_2\rangle-\tfrac32|\langle\lambda-\lambda'\mid\a_2\rangle | , 3|\langle\lambda-\lambda'\mid\a_2\rangle|)
\\
\nonumber&&\quad\cup  k_4 O\left(\min\left\{\langle \lambda\mid2\a_1+3\a_2\rangle +\langle\lambda'\mid\a_1\rangle,|\langle\lambda-\lambda'\mid\a_1\rangle-3\langle\lambda'\mid\a_2\rangle|\right\}\right.,
\\
\nonumber&&\quad\qquad\quad\left.3\min\left\{\langle\lambda\mid\a_2\rangle+\langle\lambda'\mid\a_1+2\a_2\rangle,
|\langle\lambda\mid\a_1+\a_2\rangle-\langle\lambda'\mid\a_2\rangle|\right\}\right)
\\
\nonumber&&\quad\cup k_5 O\left(\min\left\{\langle\lambda\mid\a_1\rangle+\langle\lambda'\mid2\a_1+3\a_2\rangle,\,
|\langle\lambda-\lambda'\mid\a_1\rangle+ 3\langle\lambda\mid\a_2\rangle|\right\}\right.,
\\
\nonumber&&\quad\quad\quad\left. 3
\min\{\langle\lambda\mid\a_1+2\a_2\rangle+\langle\lambda'\mid\a_2\rangle,|\langle\lambda\mid\a_2\rangle- \langle\lambda'\mid\a_1+\a_2\rangle|\}\right)
\\
\nonumber&&\quad\cup k_6 O\left(\min\{|\langle\lambda\mid2\a_1+3\a_2\rangle-\langle\lambda'\mid\a_1\rangle|,|\langle\lambda\mid\a_1\rangle -\langle\lambda'\mid2\a_1+3\a_2\rangle|\},\right.
\\
\nonumber&&\quad\quad \quad 3\min\{\langle\lambda\mid\a_1+2\a_2\rangle+\langle \lambda'\mid\a_2\rangle,
\\
\nonumber&&\quad\quad\quad\quad\langle\lambda\mid\a_2\rangle+\langle\lambda'\mid\a_1+2\a_2 \rangle
,|\langle\lambda-\lambda'\mid\a_1+\a_2\rangle|\})
\\
\nonumber&&\quad\cup k_7 O(\min\{|\langle\lambda-\lambda'\mid\a_1\rangle|,|\langle \lambda-\lambda'\mid\a_1+3\a_2\rangle|,|\langle\lambda-\lambda'\mid2\a_1+3\a_2\rangle|\},
\\
\nonumber&&\quad\qquad\quad 3\min\{|\langle \lambda-\lambda'\mid\a_2\rangle|,|\langle\lambda-\lambda'\mid\a_1+\a_2\rangle|, |\langle\lambda-\lambda'\mid\a_1+2\a_2\rangle|\})
\\
\nonumber&&\quad\cup k_8 O(\min\{\langle\lambda+\lambda'\mid\a_1\rangle,|\langle\lambda-\lambda'\mid \a_1+3\a_2\rangle+\langle \lambda\mid\a_1\rangle|,
\\
\nonumber&&\quad\quad|\langle\lambda-\lambda'\mid\a_1+3\a_2\rangle-\langle \lambda'\mid\a_1\rangle|\},
 3\min\{|\langle\lambda\mid \a_1+\a_2\rangle-\langle\lambda'\mid\a_2\mid\rangle|,\\
\nonumber&&\quad\quad\quad\quad
|\langle\lambda-\lambda'\mid \a_1+2\a_2\rangle|,|\langle\lambda\mid\a_2\rangle-\langle\lambda'\mid\a_1+\a_2\rangle|\})
\\
\nonumber&&\quad\cup k_9 O(\min\{|\langle\lambda-\lambda'\mid\a_1\rangle+3\langle\lambda\mid\a_2\rangle|, |\langle\lambda-\lambda'\mid2\a_1+3\a_2\rangle |,
\\
\nonumber&&\quad\quad  \quad
|\langle\lambda-\lambda'\mid\a_1\rangle-3\langle\lambda'\mid\a_2\rangle|\},
 3\min\{\langle\lambda+\lambda'\mid\a_2\rangle,\\
\nonumber&&\quad\quad\quad|\langle\lambda-\lambda'\mid\a_1+\a_2\rangle+\langle\lambda\mid\a_2\rangle|, |\langle\lambda-\lambda'\mid\a_1+\a_2\rangle-\langle\lambda'\mid\a_2\rangle|\})
\\
\nonumber&&\quad\cup k_{10} O(\min\{\langle\lambda+\lambda'\mid\a_1\rangle+3\langle\lambda'\mid\a_2\rangle,
| \langle\lambda\mid 2\a_1+3\a_2\rangle-\langle\lambda'\mid\a_1\rangle|,\\
\nonumber&&\quad\quad     \quad
|\langle\lambda-\lambda'\mid\a_1+3\a_2\rangle -\langle\lambda'\mid\a_1\rangle|\},
 3\min\{\langle\lambda\mid \a_1+\a_2\rangle+\langle\lambda'\mid\a_2\rangle,
 \\
\nonumber&&\quad\quad\quad  \quad
|\langle\lambda-\lambda'\mid\a_1+2\a_2\rangle+\langle \lambda'\mid\a_2\rangle|,|\langle\lambda\mid\a_2\rangle-\langle\lambda'\mid\a_1+2\a_2\rangle|\})
\\
\nonumber&&\quad\cup k_{11} O(\min\{\langle\lambda\mid\a_1+3\a_2\rangle+\langle\lambda'\mid\a_1\rangle ,|\langle\lambda-\lambda'\mid2\a_1+3\a_2\rangle+\langle\lambda'\mid\a_1\rangle|,
\\
\nonumber&&\quad\quad \quad
|\langle\lambda\mid\a_1\rangle -\langle\lambda'\mid2\a_1+3\a_2\rangle|\},
3\min\{\langle \lambda\mid\a_2\rangle+\langle\lambda'\mid\a_1+\a_2\rangle,\\
\nonumber&&\qquad\qquad\qquad|\langle\lambda-\lambda'\mid\a_1+2\a_2\rangle
-\langle\lambda\mid\a_2\rangle|,|\langle\lambda\mid\a_1+2\a_2\rangle-\langle\lambda'\mid\a_2\rangle|\})
\\
\nonumber&&\quad\cup  k_{12} O(\min\left\{\l\lambda\mid2\a_1+3\a_2\r+\l\lambda'\mid\a_1\r,
|\l\lambda-\lambda'\mid \a_1+3\a_2\r|,\right.
\\
&&\quad\quad\quad \left.\l\lambda\mid\a_1\r+\l\lambda'\mid2\a_1+3\a_2\r\right\},
\nonumber3\min\{|\l\lambda\mid\a_1+2\a_2\r-\l\lambda'\mid\a_2\r|,
\\
&&\nonumber\quad\quad\quad|\l\lambda-\lambda'\mid\a_2\r-\l\lambda'\mid\a_1+\a_2\r|\})
\end{eqnarray}
}
where the multiplicities $k_1,\dots,k_{12}$ are given in terms of orbits sizes by:
{\footnotesize
\begin{align*}
&k_1=\tfrac1{12}\frac{|O(\lambda)||O(\lambda')|}{| O(\lambda+\lambda')|}\\
&k_2=\tfrac1{12}\frac{|O(\lambda)||O(\lambda')|}{| O(|\langle\lambda-\lambda'\mid\a_1\rangle|,\tfrac32(\langle\lambda+\lambda'\mid\a_1+2\a_2\rangle- |\langle\lambda-\lambda'\mid\a_1\r|))|}\\
&k_3=\tfrac1{12}\frac{|O(\lambda)||O(\lambda')|}{|O(\langle\lambda+\lambda'\mid \a_1+\tfrac32\a_2\rangle-\tfrac32|\langle\lambda-\lambda'\mid\a_2\rangle | , 3|\langle\lambda-\lambda'\mid\a_2\rangle|)|}\\
&k_4=\tfrac1{12} |O(\lambda)||O(\lambda')|\Big/
|O\left(\min\left\{\langle \lambda\mid2\a_1+3\a_2\rangle +\langle\lambda'\mid\a_1\rangle,|\langle\lambda-\lambda'\mid\a_1\rangle-3\langle\lambda'\mid\a_2\rangle|\right\}\right.,
 \\
&\quad\quad\quad
\left.3\min\left\{\langle\lambda\mid\a_2\rangle+\langle\lambda'\mid\a_1+2\a_2\rangle,
|\langle\lambda\mid\a_1+\a_2\rangle-\langle\lambda'\mid\a_2\rangle|\right\}\right)|\\
&k_5=\tfrac1{12} |O(\lambda)||O(\lambda')|\Big/ | O\left(\min\left\{\langle\lambda\mid\a_1\rangle+\langle\lambda'\mid2\a_1+3\a_2\rangle,\,
|\langle\lambda-\lambda'\mid\a_1\rangle+ 3\langle\lambda\mid\a_2\rangle|\right\}\right.,
\\
 &\quad\quad\quad\left. 3
\min\{\langle\lambda\mid\a_1+2\a_2\rangle+\langle\lambda'\mid\a_2\rangle,|\langle\lambda\mid\a_2\rangle- \langle\lambda'\mid\a_1+\a_2\rangle|\}\right)|\\
&k_6=\tfrac1{12} |O(\lambda)||O(\lambda')|\Big/|O\left(\min\{|\langle\lambda\mid2\a_1+3\a_2
\rangle-\langle\lambda'\mid\a_1\rangle|,|\langle\lambda\mid\a_1\rangle -\langle\lambda'\mid2\a_1+3\a_2\rangle|\},\right.
\\
&\quad\quad \quad 3\min\{\langle\lambda\mid\a_1+2\a_2\rangle+\langle \lambda'\mid\a_2\rangle,
\langle\lambda\mid\a_2\rangle+\langle\lambda'\mid\a_1+2\a_2 \rangle
,|\langle\lambda-\lambda'\mid\a_1+\a_2\rangle|\})|\\
&k_7=\tfrac1{12}|O(\lambda)||O(\lambda')|\Big/|O(\min\{|\langle\lambda-\lambda'\mid\a_1\rangle|,|\langle \lambda-\lambda'\mid\a_1+3\a_2\rangle|,|\langle\lambda-\lambda'\mid2\a_1+3\a_2\rangle|\},
\\
&\quad\quad\quad 3\min\{|\langle \lambda-\lambda'\mid\a_2\rangle|,|\langle\lambda-\lambda'\mid\a_1+\a_2\rangle|, |\langle\lambda-\lambda'\mid\a_1+2\a_2\rangle|\})|\\
&k_8=\tfrac1{12} |O(\lambda)||\lambda')|\Big/|O(\min\{\langle\lambda+\lambda'\mid\a_1\rangle,|\langle\lambda-\lambda'\mid \a_1+3\a_2\rangle+\langle \lambda\mid\a_1\rangle|,
\\
 &\quad\quad\quad|\langle\lambda-\lambda'\mid\a_1+3\a_2\rangle-\langle \lambda'\mid\a_1\rangle|\},
 3\min\{|\langle\lambda\mid \a_1+\a_2\rangle-\langle\lambda'\mid\a_2\mid\rangle|,\\
 &\quad\quad\quad\quad
|\langle\lambda-\lambda'\mid \a_1+2\a_2\rangle|,|\langle\lambda\mid\a_2\rangle-\langle\lambda'\mid\a_1+\a_2\rangle|\})|\\
&k_9=\tfrac1{12} |O(\lambda)||O(\lambda')|\Big/|O(\min\{|\langle\lambda-\lambda'\mid\a_1\rangle+3\langle\lambda\mid\a_2\rangle|, |\langle\lambda-\lambda'\mid2\a_1+3\a_2\rangle |,
\\
 &\quad\quad  \quad
|\langle\lambda-\lambda'\mid\a_1\rangle-3\langle\lambda'\mid\a_2\rangle|\},
 3\min\{\langle\lambda+\lambda'\mid\a_2\rangle,\\
 &\quad\quad\quad|\langle\lambda-\lambda'\mid\a_1+\a_2\rangle+\langle\lambda\mid\a_2\rangle|, |\langle\lambda-\lambda'\mid\a_1+\a_2\rangle-\langle\lambda'\mid\a_2\rangle|\})|\\
&k_{10}=\tfrac1{12} |O(\lambda)||O(\lambda')|\Big/|O(\min\{\langle\lambda+\lambda'\mid\a_1\rangle+3\langle\lambda'\mid\a_2\rangle,
| \langle\lambda\mid 2\a_1+3\a_2\rangle-\langle\lambda'\mid\a_1\rangle|,\\
 &\quad\quad     \quad
|\langle\lambda-\lambda'\mid\a_1+3\a_2\rangle -\langle\lambda'\mid\a_1\rangle|\},
 3\min\{\langle\lambda\mid \a_1+\a_2\rangle+\langle\lambda'\mid\a_2\rangle,
 \\
 &\quad\quad\quad  \quad
|\langle\lambda-\lambda'\mid\a_1+2\a_2\rangle+\langle \lambda'\mid\a_2\rangle|,|\langle\lambda\mid\a_2\rangle-\langle\lambda'\mid\a_1+2\a_2\rangle|\})|\\
&k_{11}=\tfrac1{12} |O(\lambda)||O(\lambda')|\Big/|O(\min\{\langle\lambda\mid\a_1+3\a_2\rangle+\langle\lambda'\mid\a_1\rangle ,|\langle\lambda-\lambda'\mid2\a_1+3\a_2\rangle+\langle\lambda'\mid\a_1\rangle|,
\\
 &\quad\quad \quad
|\langle\lambda\mid\a_1\rangle -\langle\lambda'\mid2\a_1+3\a_2\rangle|\},
3\min\{\langle \lambda\mid\a_2\rangle+\langle\lambda'\mid\a_1+\a_2\rangle,\\
 &\quad\quad|\langle\lambda-\lambda'\mid\a_1+2\a_2\rangle
-\langle\lambda\mid\a_2\rangle|,|\langle\lambda\mid\a_1+2\a_2\rangle-\langle\lambda'\mid\a_2\rangle|\})|\\
&k_{12}=\tfrac1{12} |O(\lambda)||O(\lambda')|\Big/|O(\min\left\{\l\lambda\mid2\a_1+3\a_2\r+\l\lambda'\mid\a_1\r,
|\l\lambda-\lambda'\mid \a_1+3\a_2\r|,\right.
\\
 &\quad\quad\quad \left.\l\lambda\mid\a_1\r+\l\lambda'\mid2\a_1+3\a_2\r\right\},
 3\min\{|\l\lambda\mid\a_1+2\a_2\r-\l\lambda'\mid\a_2\r|,
\\
 &\quad\quad\quad\quad|\l\lambda-\lambda'\mid\a_2\r-\l\lambda'\mid\a_1+\a_2\r|\})|\;.
\end{align*}
}
\end{prop}

\medskip

As earlier the prove of \eqref{G2product} could be done by writing down the existing list of all special cases following from it. Then each case of the list is easily verify.

\begin{example}\
\begin{equation}\label{G2example}
\begin{alignedat}{1}
&O(1,2)\otimes O(3,1)=O(4, 3) \cup O(5, 1)\cup  O(3, 4) \cup O(2, 6)\cup O(1, 6) \cup O(1, 5)\\
&\qquad\cup O(1, 3) \cup 2 O(0, 8)\cup 2 O(0, 6)\cup 2 O(0, 4)\cup 2O(0, 3)\cup O(1, 1).
\end{alignedat}
\end{equation}
Orbits and decomposition of  product \eqref{G2example} are presented in figure \ref{rysG2}. More advanced examples one could find in appendix C.
\begin{figure}[h]
\begin{center}
\subfigure[][]{\label{rysG2a}\includegraphics[scale=0.32]{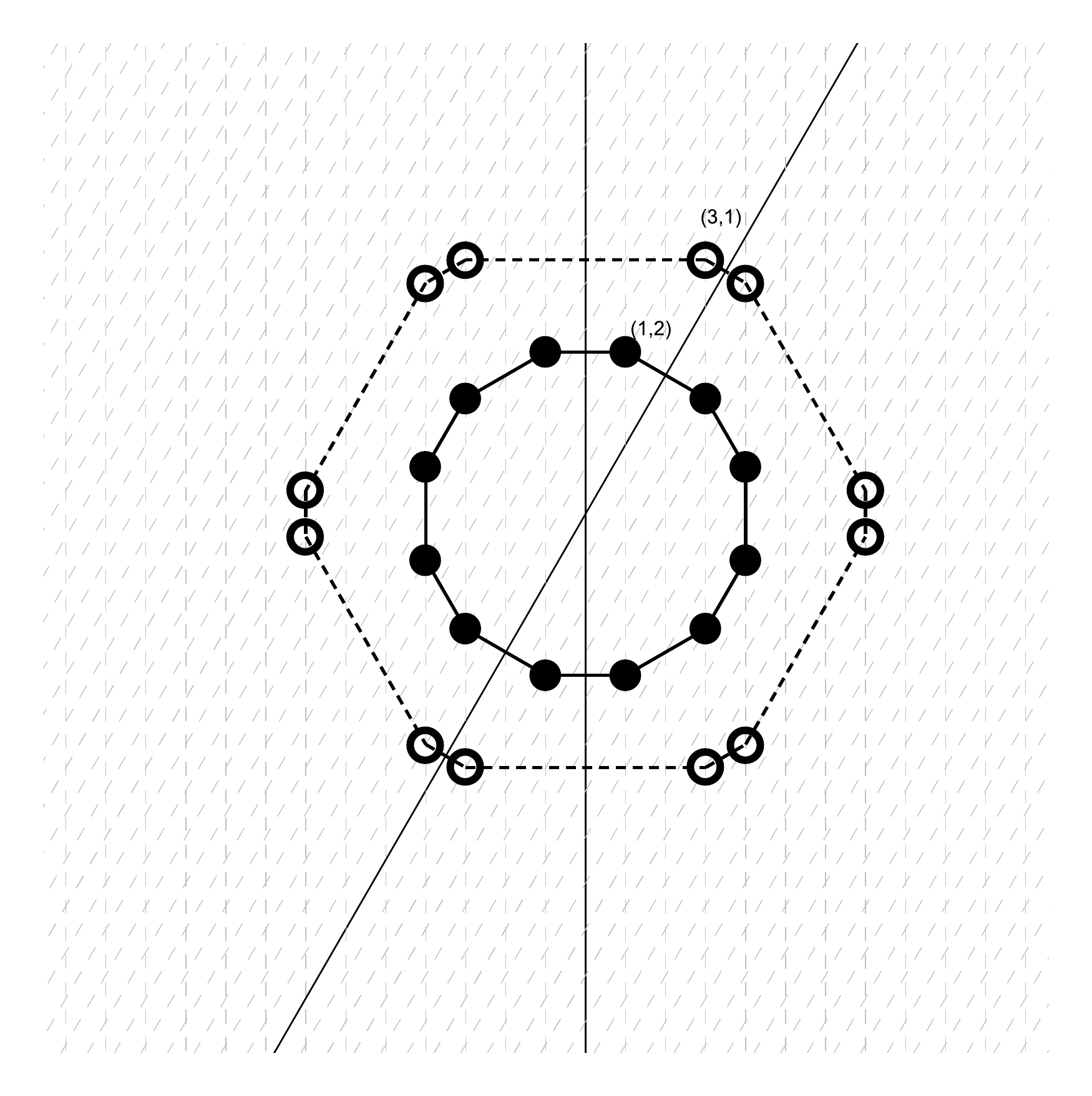}}
 \;
\subfigure[][]{\label{rysG2b}\includegraphics[scale=0.32]{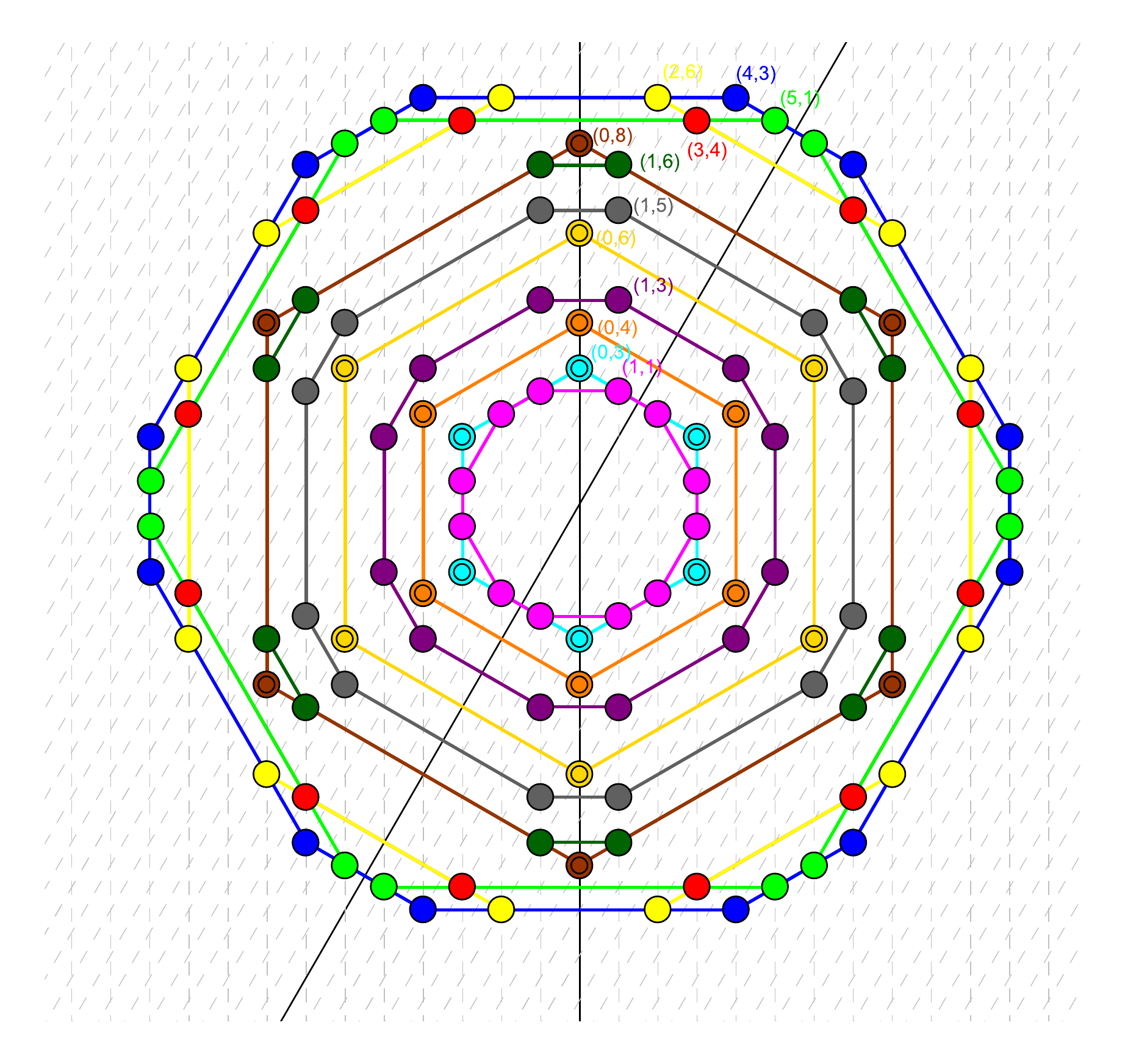}}
\caption{\subref{rysG2a} The twelve points of the orbit $O(1,2)$ and of $O(3,1)$ of $G_2$. The straight lines are the reflection mirrors containing $\omega_1$ and $\omega_2$. The dominant points are indicated in the positive sector. Dashed lines are the directions of the weight lattice axes of $G_2$.\newline
\subref{rysG2b} The twelve orbits of the decomposition of the product $O(1,2)\otimes O(3,1)$ of $G_2$. Eight of them are dodecagons    and four are hexagons taken twice what is denoted by double dots. The straight lines are the reflection mirrors containing $\omega_1$ and $\omega_2$. The dominant points are indicated in the positive sector. Dashed lines are the directions of the weight lattice axes of $G_2$.}\label{rysG2}
\end{center}
\end{figure}\
Using \eqref{granice} one can get
\begin{equation*}
\begin{alignedat}{1}
& \lim_{x\rightarrow \tfrac12}O(1,2)\otimes O(x,1)\\
&\qquad  =O(\tfrac32, 3)\cup O(\tfrac52, 1)\cup O(\tfrac12, 4) \cup O(\tfrac12, \tfrac92)\cup O(\tfrac32, \tfrac32)\cup O(1, \tfrac52)\cup
 2 O(1, \tfrac32)\\
 &\qquad\qquad\cup O(\tfrac52, \tfrac12) \cup O(1, \tfrac12)\cup O(\tfrac12, \tfrac32)
  \cup O(\tfrac12, 1)
\\
&\qquad  =O(1,2)\otimes O(\tfrac12,1).
\end{alignedat}
\end{equation*}

\end{example}

%%%%%%%%%%%%%%%%%%%%%%%%%%%%%%%%%%%%%%%%%%%
\section{Decompositions of products of orbits of  $H_2$}

Last group considered in the paper is non-crystallographic group $H_2$.
Some basic information about that group  are presented in Appendix D.

A useful general symmetry property of all orbits of $H_2$:
\begin{gather}\label{autoH}
\lambda\in O(a,b)\quad \Longleftrightarrow\quad -\lambda\in O(b,a)\,,\qquad a,b\in\R\,.
\end{gather}
In most cases we are interested in $a,b\in\Z[\tau]$, however \eqref{autoH} is valid for any real $a$~and~$b$.

For this group one has also three kinds of nontrivial orbits. One can find it for example in \cite{HLP08}:
\begin{alignat*}{2}
O(0,0)    &=\{(0,0)\},\\
O(a, 0)   &= \{(a, 0), (-a, a\tau), (a\tau,-a\tau), (-a\tau, a), (0,-a)\},\\
O(0, b) &= \{(0, b), (b\tau,-b), (-b\tau, b\tau), (b,-b\tau), (-b, 0)\},\\
\\
O(a, b) &= \left\{(a, b), (-a, b + a\tau ), (a\tau + b\tau,-b - a\tau),\right.\\
          &  (-a\tau - b\tau, a + b\tau ), (b,-a - b\tau ), (a + b\tau,-b),\\
          &  \left.(-a- b\tau, a\tau + b\tau ), (b + a\tau,-a\tau - b\tau),(-b-a\tau, a), (-b,-a)\right\}.
\end{alignat*}

As before product of two orbits of $H_2$ is written in terms of $\lambda=(a_1,a_2)=a_1\omega_1+a_2\omega_2$ and $\lambda'=(b_1,b_2)=b_1\omega_1+b_2\omega_2$ and simple roots, i.e.:
\begin{equation*}
\begin{alignedat}{2}
\langle\lambda\mid\alpha_j\rangle
& =a_j, \\
\langle\lambda'\mid\alpha_j\rangle
& =b_j,\\
\end{alignedat}
\qquad\textrm{ for }j=1,2\textrm{ and }a_1,a_2,b_1,b_2\in\R^{\geq0}.
\end{equation*}

\begin{prop}\label{H2decomposition}\

Decomposition of the product \eqref{decomposition} of two  orbits of $H_2$  with dominant weights $\lambda=(a_1,a_2)$ and $\lambda'=(b_1,b_2)$  for $a_1,a_2,b_1,b_2\in\R^{\geq0}$  is given by the following formula
%%%%%%%%%%%%%%%%%%%%%%%%%%%%%%%%%%%%%
{%\tiny
\begin{eqnarray}\label{H2product}
&&O(\lambda)\otimes O(\lambda')
 \\
\nonumber&&\qquad = k_1 O(\lambda+\lambda')% \qquad\qquad\qquad\qquad\qquad\qquad a_1,a_2,b_1,b_2\in\Z^{\geq0}[\tau]
\\
\nonumber&&\qquad\cup   k_2 O(|\langle\lambda-\lambda'\mid\a_1\rangle |, \langle\lambda+\lambda'\mid\tfrac\tau2\a_1+\a_2\rangle-\tfrac\tau2|\langle\lambda-\lambda'\mid\a_1\rangle|)
\\
\nonumber&&\qquad\cup k_3  O(\min\{\langle\lambda\mid\a_1\rangle+\tau\langle\lambda'\mid\a_1+\a_2\rangle, |\langle\lambda-\lambda'\mid\a_1\rangle+\tau\langle \lambda\mid\a_2\rangle|\},
\\
\nonumber&&\qquad\qquad
\min\{\langle\lambda'\mid\a_2\rangle+\tau\langle\lambda\mid\a_1+\a_2\rangle, |\langle\lambda-\lambda'\mid\a_2\rangle-\tau\langle \lambda'\mid\a_1\rangle|\} )
\\
\nonumber&&\qquad\cup  k_4 O(\min\{|\langle\lambda\mid\a_1\rangle-\tau\langle\lambda'\mid\a_1+\a_2\rangle|,
|\langle\lambda'\mid\a_1\rangle-\tau\langle\lambda\mid\a_1+\a_2\rangle|\},
\\
\nonumber&&\qquad\qquad
\min\{\langle \lambda\mid\a_2\rangle+\langle\lambda'\mid\a_1+\tau\a_2\rangle,
\langle\lambda'\mid\a_2\rangle+\langle \lambda\mid\a_1+\tau\a_2\rangle,
\\
\nonumber&&\qquad\qquad\qquad
|\langle\lambda-\lambda'\mid\tau\a_1+\a_2\rangle|\})
\\
\nonumber&&\qquad\cup k_5 O( \left|\min\left\{\langle\lambda\mid\tau\a_1+\a_2\rangle-\langle\lambda'\mid\a_1\rangle, \langle\lambda\mid\a_1\rangle +\langle\lambda'\mid\a_2\rangle,\right.\right.
\\
\nonumber&&\qquad\qquad
\left.\left.|\langle \lambda\mid\a_1+\tau\a_2\rangle-\tau\langle \lambda'\mid\a_1+\a_2\rangle|\right\}\right|,
\left|\min\{\langle\lambda\mid\a_1\rangle +\langle\lambda'\mid\a_2\rangle,
\right.
\\
\nonumber&&\qquad\qquad
\left.
\langle\lambda'\mid\a_1+\tau\a_2\rangle-\langle\lambda\mid\a_2\rangle,
|\langle\lambda'\mid\tau\a_1+\a_2\rangle-\tau\langle\lambda\mid\a_1+\a_2\rangle|\}\right|)
\\
\nonumber&&\qquad\cup k_6 O(\langle \lambda+\lambda'\mid\a_1+\tfrac\tau2\a_2\rangle-\tfrac\tau2|\langle\lambda-\lambda'\mid\a_2\rangle|\}, |\langle\lambda-\lambda'\mid\a_2\rangle|)
\\
\nonumber&&\qquad\cup k_7 O(\left|\min\{\l\lambda'-\lambda\mid\a_1\r+\tau\l\lambda'\mid\a_2\r,\l\lambda'\mid\a_1\r+
\tau\l\lambda\mid\a_1+\a_2\r\}\right|,
\\
\nonumber&&\qquad\qquad\left|\min\{\l\lambda\mid\a_2\r+\tau\l\lambda'\mid\a_1+\a_2\r,
\l\lambda-\lambda'\mid\a_2\r+\tau\l\lambda\mid\a_1\r\}\right|)
\\
\nonumber&&\qquad\cup k_8 O(\min\{\l\lambda\mid\a_1\tau +\a_2\r + \l\lambda'\mid\a_1\r ,
 \l\lambda\mid\a_1\r +\l\lambda'\mid\a_1\tau+\a_2\r,
 \\
 \nonumber&&\qquad\qquad|\l\lambda-\lambda'\mid\a_1+\tau\a_2\r)|\},
  \min\{ |\l\lambda'\mid\a_2\r-\tau\l\lambda\mid\a_1+\a_2\r|,
  \\
   \nonumber&&\qquad\qquad\qquad
  |\l\lambda\mid\a_2\r-\tau\l\lambda'\mid\a_1+\a_2\r)|\})
\\
\nonumber&&\qquad\cup k_9 O(\left|\min\{-\l\lambda\mid\a_1\mid\r+\l\lambda'\mid\tau\a_1+\a_2\r,
\l\lambda\mid\a_2\r+\l\lambda'\mid\a_1\r,\right.
\\
 \nonumber&&\qquad\qquad\qquad
\left.|\l\lambda'\mid\a_1+\tau\a_2\r-\tau\l\lambda\mid\a_1+\a_2\r|\}\right|,
\\
 \nonumber&&\qquad\qquad\left|\min\left\{\l\lambda\mid\a_1+\tau\a_2\r-\l\lambda'\mid\a_2\r,
 \l\lambda\mid\a_2\r+\l\lambda'\mid\a_1\r,\right.\right.
 \\
 \nonumber&&\qquad\qquad\qquad\left.\left.
 |\l\lambda\mid\tau\a_1+\a_2\r-\tau\l\lambda'\mid\a_1+\a_2\r|\right\}\right|)
\\
\nonumber&&\qquad\cup k_{10} O\left(A,B\right)\;,
\end{eqnarray}
}
where the multiplicities $k_1,\dots,k_{10}$ are given in terms of orbits sizes by:
\begin{eqnarray*}
  A&=&\Big| | \tau \max\{  \l\lambda\mid\a_2\r - \l\lambda'\mid\a_1\r,  \l\lambda'\mid\a_2\r-\l\lambda\mid\a_1\r \}
\\
\nonumber&&\!\!\!\!\!\!\!\!
  \cdot\Big(1 - \mathrm{sign}\left(\max\{0, \l\lambda\mid\a_2\r - \l\lambda'\mid\a_1\r,
  \l\lambda'\mid\a_2\r -\l\lambda\mid\a_1\r \}\right)\Big)|
  \\
  \nonumber&&\!\!\!\!\!\!\!\!
 + \min\{\l\lambda\mid\a_2\r - \l\lambda'\mid\a_1\r,
  \l\lambda'\mid\a_2\r-\l\lambda\mid\a_1\r  ,
  \tau |\l\lambda-\lambda'\mid\a_1+\a_2\r |\}\Big|,
  \\
B&=&\Big|
\Big(\l\lambda\mid\tau\a_1+(1+\tau)\a_2 \r - \l\lambda'\mid(1+\tau)\a_1+\tau\a_2\r \Big)
\\
\nonumber&&
\!\!\!\!\!\!\!\!\!\!\cdot
\Big(1 -|\mathrm{sign}\left(\min\{0, \tau \l\lambda-\lambda'\mid\a_1+\a_2\r,
  \l\lambda'\mid\tau\a_1+\a_2\r-\l\lambda\mid\a_1+\tau\a_2\r  \}\right)|\Big)
  \\
\nonumber&&\!\!\!\!\!\!\!\!\!\!
    +\Big(1 - |\mathrm{sign}\left(\min\{0, \l\lambda\mid\a_2\r - \l\lambda'\mid\a_1\r,
 \l\lambda'\mid\a_2\r-\l\lambda\mid\a_1\r \}\right)|\Big)
   \\
  \nonumber&& \!
\cdot \max\{
    \l\lambda\mid\tau \a_1+\a_2\r - \l\lambda'\mid\a_1+\tau\a_2\r ,
    \l\lambda'\mid\tau\a_1+\a_2\r-\l\lambda\mid\a_1+\tau\a_2\r  \}
     \\
   \nonumber&&\!\!\!\!\!\!\!\!
    + \Big( \l\lambda\mid(1+\tau)\a_1+\tau\a_2\r -  \l\lambda'\mid\tau\a_1+(1+\tau)\a_2\r \Big)
   \\
  \nonumber&&
\cdot
    \Big(1 -
     |\mathrm{sign}\left(\min\{0, \l\lambda\mid\tau\a_1+\a_2\r - \l\lambda'\mid\a_1+\tau\a_2\r \}\right)|\Big)
     \\
 \nonumber&&\!\!\!\!\!\!\!\!\!\!\!\!\!\!\!\!\!\!\!\!\!
\cdot
     \mathrm{sign}( \max\{0, \l\lambda'\mid\tau\a_1+\a_2\r-\l\lambda\mid\a_1+\tau\a_2\r\})
     \mathrm{sign}( \min\{0, \tau\l\lambda-\lambda'\mid\a_1+\a_2\r \})
     \\
\nonumber &&\!\!\!\!\!\!\!\! +
  \max\{\l\lambda\mid\a_2\r - \l\lambda'\mid\a_1\r,  \l\lambda'\mid\a_2\r-\l\lambda\mid\a_1\r \}
     \\
  \nonumber&&\quad
\cdot
   \mathrm{sign}(\min\{0, \l\lambda\mid\a_2\r - \l\lambda'\mid\a_1\r, \l\lambda'\mid\a_2\r-\l\lambda\mid\a_1\r  \})
\Big|
  \end{eqnarray*}
and
{\footnotesize
\begin{align*}
&k_1=\tfrac1{10}\frac{|O(\lambda)||O(\lambda')|}{|O(\lambda+\lambda')|}\\
&k_2=\tfrac1{10}\frac{|O(\lambda)||O(\lambda')|}{\Big|O(|\langle\lambda-\lambda'\mid\a_1\rangle |, \langle\lambda+\lambda'\mid\tfrac\tau2\a_1+\a_2\rangle-\tfrac\tau2|\langle\lambda-\lambda'\mid\a_1\rangle|)\Big|}\\
&k_3=\tfrac1{10} |O(\lambda)||O(\lambda')|\Big/\Big| O(\min\{\langle\lambda\mid\a_1\rangle+\tau\langle\lambda'\mid\a_1+\a_2\rangle, |\langle\lambda-\lambda'\mid\a_1\rangle+\tau\langle \lambda\mid\a_2\rangle|\},
\\
 &\qquad\qquad
\min\{\langle\lambda'\mid\a_2\rangle+\tau\langle\lambda\mid\a_1+\a_2\rangle, |\langle\lambda-\lambda'\mid\a_2\rangle-\tau\langle \lambda'\mid\a_1\rangle|\} )\Big|\\
&k_4=\tfrac1{10} |O(\lambda)||O(\lambda')|\Big/\Big|
O(\min\{|\langle\lambda\mid\a_1\rangle-\tau\langle\lambda'\mid\a_1+\a_2\rangle|,
|\langle\lambda'\mid\a_1\rangle-\tau\langle\lambda\mid\a_1+\a_2\rangle|\},
\\
 &\qquad\qquad
\min\{\langle \lambda\mid\a_2\rangle+\langle\lambda'\mid\a_1+\tau\a_2\rangle,
\langle\lambda'\mid\a_2\rangle+\langle \lambda\mid\a_1+\tau\a_2\rangle,
|\langle\lambda-\lambda'\mid\tau\a_1+\a_2\rangle|\})
\Big|\\
&k_5=\tfrac1{10} |O(\lambda)||O(\lambda')|\Big/\Big|
O( \left|\min\left\{\langle\lambda\mid\tau\a_1+\a_2\rangle-\langle\lambda'\mid\a_1\rangle, \langle\lambda\mid\a_1\rangle +\langle\lambda'\mid\a_2\rangle,\right.\right.
\\
 &\qquad\qquad
\left.\left.|\langle \lambda\mid\a_1+\tau\a_2\rangle-\tau\langle \lambda'\mid\a_1+\a_2\rangle|\right\}\right|,
\left|\min\{\langle\lambda\mid\a_1\rangle +\langle\lambda'\mid\a_2\rangle,
\right.
\\
 &\qquad\qquad\qquad
\left.
\langle\lambda'\mid\a_1+\tau\a_2\rangle-\langle\lambda\mid\a_2\rangle,
|\langle\lambda'\mid\tau\a_1+\a_2\rangle-\tau\langle\lambda\mid\a_1+\a_2\rangle|\}\right|)
\Big|\\
&k_6=\tfrac1{10}\frac{|O(\lambda)||O(\lambda')|}{|O(\langle \lambda+\lambda'\mid\a_1+\tfrac\tau2\a_2\rangle-\tfrac\tau2|\langle\lambda-\lambda'\mid\a_2\rangle|\}, |\langle\lambda-\lambda'\mid\a_2\rangle|)|}\\
&k_7=\tfrac1{10} |O(\lambda)||O(\lambda')\Big/|O(\left|\min\{\l\lambda'-\lambda\mid\a_1\r+\tau\l\lambda'\mid\a_2\r,\l\lambda'\mid\a_1\r+
\tau\l\lambda\mid\a_1+\a_2\r\}\right|,
\\
&\qquad\qquad\left|\min\{\l\lambda\mid\a_2\r+\tau\l\lambda'\mid\a_1+\a_2\r,
\l\lambda-\lambda'\mid\a_2\r+\tau\l\lambda\mid\a_1\r\}\right|)|\\
&k_8=\tfrac1{10} |O(\lambda)||O(\lambda')|\Big/|O(\min\{\l\lambda\mid\a_1\tau +\a_2\r + \l\lambda'\mid\a_1\r ,
 \l\lambda\mid\a_1\r +\l\lambda'\mid\a_1\tau+\a_2\r,
 \\
&\qquad\qquad|\l\lambda-\lambda'\mid\a_1+\tau\a_2\r)|\},
  \min\{ |\l\lambda'\mid\a_2\r-\tau\l\lambda\mid\a_1+\a_2\r|,
  \\
&\qquad\qquad\qquad
  |\l\lambda\mid\a_2\r-\tau\l\lambda'\mid\a_1+\a_2\r)|\})|\\
&k_9=\tfrac1{10} |O(\lambda)||O(\lambda')|
\Big/\Big|
O(\left|\min\{-\l\lambda\mid\a_1\mid\r+\l\lambda'\mid\tau\a_1+\a_2\r,
\l\lambda\mid\a_2\r+\l\lambda'\mid\a_1\r,\right.
\\
 &\qquad\qquad\qquad
\left.|\l\lambda'\mid\a_1+\tau\a_2\r-\tau\l\lambda\mid\a_1+\a_2\r|\}\right|,
\left|\min\left\{\l\lambda\mid\a_1+\tau\a_2\r-\l\lambda'\mid\a_2\r,\right.\right.
\\
 &\qquad\qquad\left.\left.
 \l\lambda\mid\a_2\r+\l\lambda'\mid\a_1\r,
 |\l\lambda\mid\tau\a_1+\a_2\r-\tau\l\lambda'\mid\a_1+\a_2\r|\right\}\right|)
\Big|\\
&k_{10}=\tfrac1{10}|\tfrac{O(\lambda)||O(\lambda')}{|O(A,B)|}\;.
\end{align*}
}
\end{prop}

The prove of \eqref{H2product} is could be again done by writing of existing list of all special cases following from it. Then each case of the list is easily verified.
\begin{example}\

\begin{equation}\label{H2example}
\begin{alignedat}{1}
&\!\!\!\! O(1,2)\otimes O(3,1)=O(4, 3)\cup O(2, 3 + \tau)\cup  O(2 - \tau,  3 \tau-1)
\cup O(\tau-1 , 1) \\
&\quad \cup O(2 + \tau, 1 + \tau)  \cup O(4 + \tau, 1)
\cup O( 2 \tau-3 ,  3 \tau -2)\cup O(2 \tau-2 , 3 \tau-1 )  \\
 &\quad  \cup O(3 \tau-3 ,  2 \tau-1 )\cup 2 O(1, 0).
\end{alignedat}
\end{equation}
Orbits and decomposition of  product \eqref{H2example} are presented in figure \ref{rysH2}. More advanced examples one could find in appendix D.
\begin{figure}[h!]
\begin{center}
\subfigure[][]{\label{rysH2a}\includegraphics[scale=0.37]{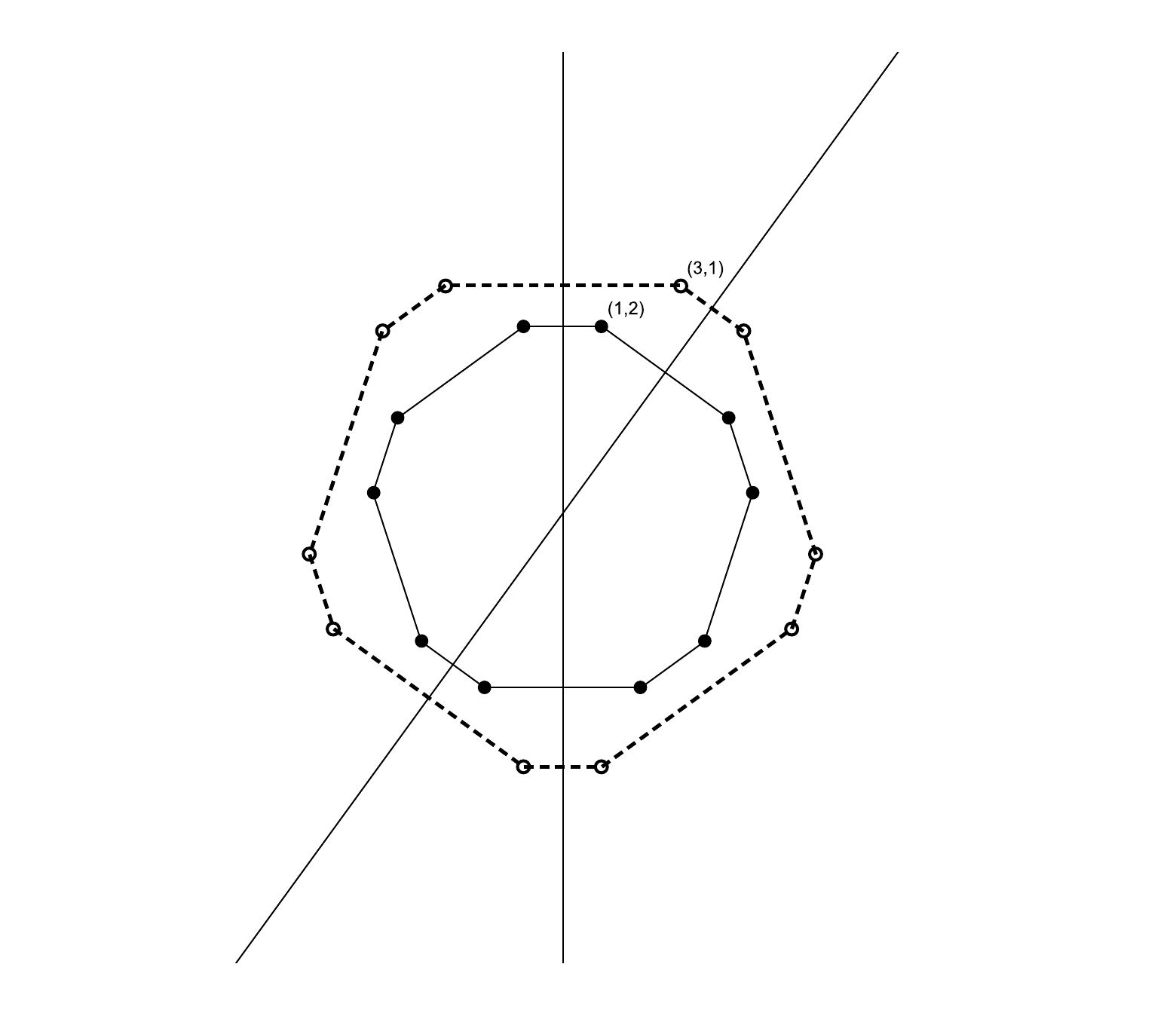}}
\;
\subfigure[][]{\label{rysH2b}\includegraphics[scale=0.37]{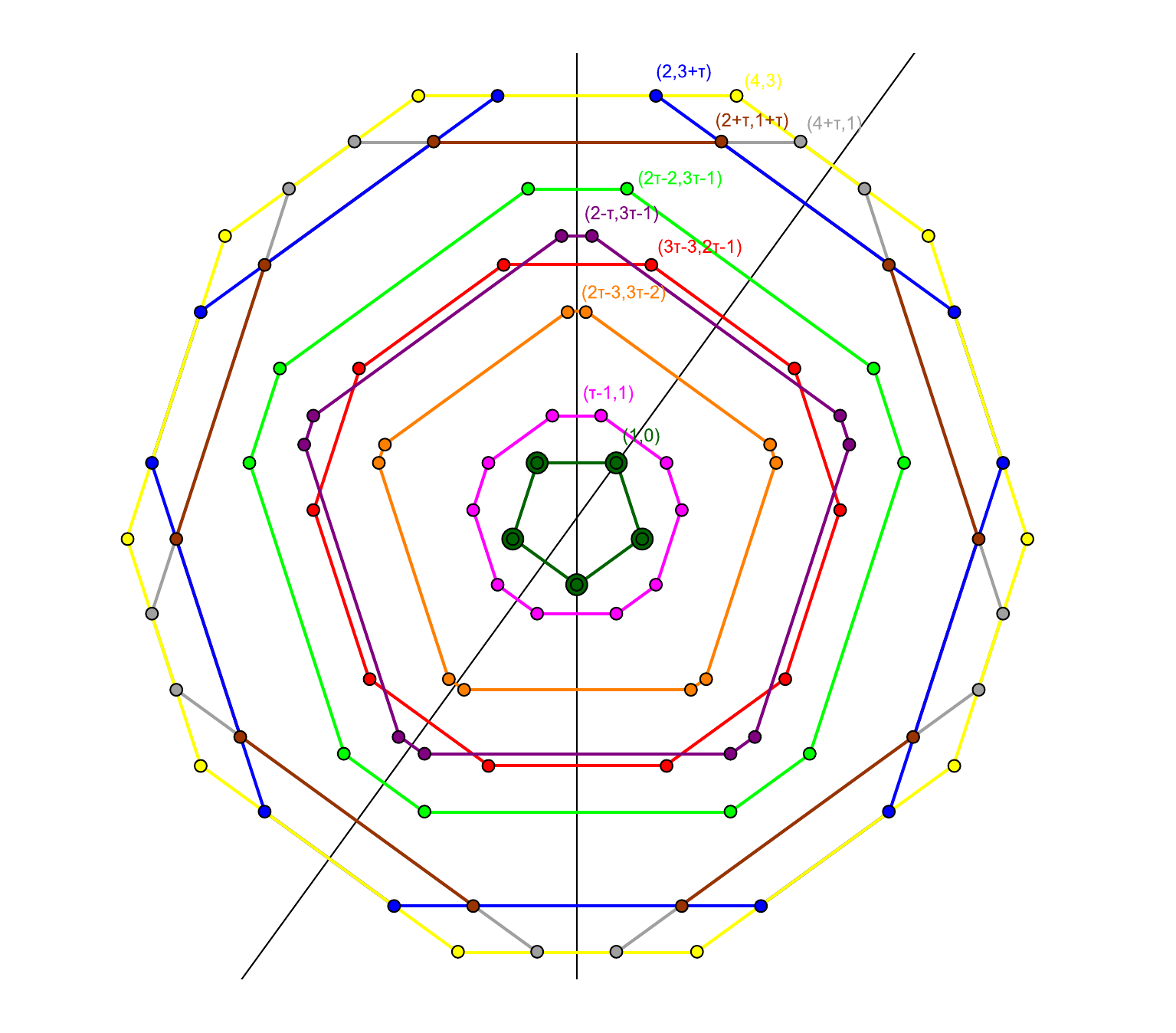}}
\caption{\subref{rysH2a}
The ten points of the orbit $O(1,2)$ and of $O(3,1)$ of $H_2$. The straight lines are the reflection mirrors containing $\omega_1$ and $\omega_2$. The dominant points are indicated in the positive sector.\newline
\subref{rysH2b}
The ten orbits of the decomposition of the product $O(1,2)\otimes O(3,1)$ of $H_2$. Nine of them are decagons    and one  is pentagon  taken twice what is denoted by double dots. The straight lines are the reflection mirrors containing $\omega_1$ and $\omega_2$. The dominant points are indicated in the positive sector.
}\label{rysH2}
\end{center}
\end{figure}\
Using \eqref{granice} one can get
\begin{equation*}
\begin{alignedat}{1}
& \lim_{y\rightarrow \tau}O(1,y)\otimes O(3,1)\\
&\quad  = O(4, 1+\tau)\cup
O(1, 2\tau) \cup O(2, 1+2\tau)\cup O(2-\tau,1+\tau)\cup O(\tau-1, 1+2\tau)
\\
&\qquad \cup O(2+\tau, 2\tau-1)\cup O(4+\tau,\tau-1)\cup O(2\tau-2,1+\tau)\\
&\qquad \cup O(2\tau-3,\tau)\cup 2O(3-\tau, 0)\\
&\quad  =O(1,\tau)\otimes O(3,1).
\end{alignedat}
\end{equation*}

\end{example}

\section*{Remarks}

The orbits of Weyl groups of simple algebra are part of the weight system of irreducible representations. Working with the representations it can be advantageous  to work with simple objects, namely  the Weyl group orbits rather then the weight systems. The group $H_2$ is in the  physics the dihedral group  of order $10$ and plays an important role in modeling two dimensional quasicrystals, see \cite{T01, PT02, GPMD09}.

Orbits of the group are indispensable in defining families of orthogonal functions and polynomials \cite{KP06,KP07,KP08}. The discretization of the orbit functions is interesting problem known for the Weyl groups \cite{MP87} but is not known for $H_2$ functions  and polynomials, which are rather promising for exploration in digital data processing. The orthogonality of orbit functions of group $H_2$ has not been found yet and it is challenging problem to be solved.

Similar problems for $H_3$ and $H_4$ groups could be solved by the same way.

The orbits could be viewed as polytops. The decomposition of their product can be seen as an onion-like structure form by concentric orbits. Such types of structure are very interesting for example to describe carbon and carbon nanotubes, see \cite{HLP08,T}.

The formulas \eqref{A2product},\eqref{C2product},\eqref{G2product}\eqref{H2product} work also in the case when the orbits are points not belonging to the weight latices. Their coordinates of their dominants weights are nonnegative real numbers.   The congruence classes are lost in this case. Illustration of these is in examples of $G_2$ and $H_2$, and limits were calculated for such values. This fact suggest that orthogonality of orbit functions for nonnegative real numbers should work too. This is good starting  point to another paper.

Calculation of any product of two orbits of considered groups  is specially effective when one uses computer programs as Mathematica, Maple or others. For such calculations it is enough to use elementary functions. One could also found, by analogy, formulae for other groups not described in the paper.

\subsection*{Acknowledgements}\

The author  expresses her  gratitude to Centre de Recherches  Math\'ematiques, Universit\'e de Montr\'eal   for the hospitality extended to her during her postdoctoral fellowship and to  MITACS and ODDA Technologies for partial support of this work.
 The author is also grateful to J.Patera for helpful comments and advice.

%%%%%%%%%%%%%%%%%%%%%%%%%%%%%%%%%%%%%%%%%%%%%%%%%%%%%%%%%%%%%%%%%%%%%%%%%%%%%%%%%%%%%%%%%

%\newpage
\appendix
%\addappheadtotoc
%\appendixpage
\appendixpageoff
%\chapter{}
\section{}
In this section sketch of proof of proposition \ref{A2decomposition} is presented.

Firstly when one rewrite \eqref{A2product} in terms of  \eqref{zmienneA2} then gets
\begin{equation}
\begin{aligned}\label{A2_1product}
&O(a_1,a_2)\otimes O(b_1,b_2)
 \\
&\qquad = k_1\;O(a_1+b_1,a_2+b_2)
\qquad\qquad\qquad\qquad\qquad a_1,a_2,b_1,b_2\in\R^{\geq0}
\\
&\qquad\cup k_2\;O\left(\left|a_1-b_1 \right|,a_2+b_2+\min\{ a_1,b_1\} \right)
\\
&\qquad\cup k_3\; O\left(a_1+b_1+\min\{ a_2,b_2\}, \left|a_2-b_2 \right|\right)
\\
&\qquad\cup k_4\;O\left(\left|b_1+\min\{a_2,b_2-a_1\}\right|, \left|a_2+\min\{b_1,a_1-b_2\}\right|\right)
\\
& \qquad\cup k_5\;O\left(\left|a_1+\min\{b_2,a_2-b_1\}\right|, \left|b_2+\min\{a_1,b_1-a_2\}\right|\right)
\\
&\qquad\cup k_6 \;O(\big||a_1+a_2-b_1-b_2|-|\min\{a_1-b_2,b_1-a_2,0\}|\big|,
\\
&\qquad\qquad\qquad \big||a_1+a_2-b_1-b_2|-|\min\{-a_1+b_2,-b_1+a_2,0\}|\big|),
 \end{aligned}
 \end{equation}
where
{ \begin{align*}
&k_1=\! \tfrac16\tfrac{|O(a_1,a_2)||O(b_1,b_2)|}{|O(a_1+b_1,a_2+b_2)|}
\\
&k_2=\! \tfrac16\tfrac{|O(a_1,a_2)||O(b_1,b_2)|}
     {|O\left(\left|a_1-b_1 \right|,a_2+b_2+\min\{ a_1,b_1\} \right)|}
\\
&k_3=\!\tfrac16\tfrac{|O(a_1,a_2)||O(b_1,b_2)|}{|O\left(a_1+b_1+\min\{ a_2,b_2\}, \left|a_2-b_2 \right|\right)|}\\
&k_4= \tfrac16\tfrac{|O(a_1,a_2)||O(b_1,b_2)|}{|O\left(\left|b_1+\min\{a_2,b_2-a_1\}\right|, \left|a_2+\min\{b_1,a_1-b_2\}\right|\right)|}\\
&k_5=\! \tfrac16\tfrac{|O(a_1,a_2)||O(b_1,b_2)|}{|O\left(\left|a_1+\min\{b_2,a_2-b_1\}\right|, \left|b_2+\min\{a_1,b_1-a_2\}\right|\right)|}\\
&k_6=\! \tfrac{\!|O(a_1,a_2)||O(b_1,b_2)|}{6\big|O\big(\big||a_1+a_2-b_1-b_2|-|\min\{a_1-b_2,b_1-a_2,0\}|\big|,
\big||a_1+a_2-b_1-b_2|-|\min\{-a_1+b_2,-b_1+a_2,0\}|\big|\big)\big|}.
\end{align*}
}
It is obvious that using \eqref{zmienneA2} one gets from \eqref{A2product} equation \eqref{A2_1product}.

\begin{pf}
 Now to check \eqref{A2_1product} it is enough to consider all special cases for $a_1,a_2,b_1,b_2\in\R^{\geq0}.$
\begin{itemize}
 \item
 First, let consider $a_1=a_2=b_1=b_2=0$, then
 \begin{align*}
&  O(0,0)\otimes O(0,0)\\
&\qquad=\tfrac16O(0,0)\cup\tfrac16O(0,0)\cup \tfrac16O(0,0)\cup\tfrac16O(0,0)
\cup\tfrac16O(0,0)\cup\tfrac16O(0,0)
 \\
 &\qquad=O(0,0).
  \end{align*}
 Similarly, when $a_1,a_2\neq 0$ and $b_1=b_2=0$, then
\begin{align*}
&O(a_1,a_2)\otimes O(0,0) \\ &\quad=\tfrac16O(a_1,a_2)\cup\tfrac16O(a_1,a_2)\cup\tfrac16O(a_1,a_2) \cup\tfrac16O(a_1,a_2)\cup\tfrac16O(a_1,a_2)
\\&\quad
\cup\tfrac16O(a_1,a_2) =O(a_1,a_2).
\end{align*}

\item
Next special case which one gets from \eqref{A2product} or equivalently from \eqref{A2_1product} for $0\neq a_1\neq b_1\neq 0$ and $a_2=b_2=0$ following product
\begin{align*}
 & O(a_1,0)\otimes O(b_1,0)=\tfrac12O(a_1 + b_1, 0)\cup \tfrac14 O(|a_1 - b_1|, \min\{a_1, b_1\})
 \\
 &\; \cup \!\tfrac12 O(a_1+b_1,0)\cup\!\tfrac14 O(|a_1 - b_1|, \min\{a_1, b_1\})\cup \!\tfrac14 O(|a_1 - b_1|, \min\{a_1, b_1\})
 \\
 &\;=O(a_1 + b_1, 0)\cup   O(|a_1 - b_1|, \min\{a_1, b_1\})
 \cup\tfrac14 O(|a_1 - b_1|, \min\{a_1, b_1\})
\\
\nonumber
&\;=\left\{
\begin{array}{lll}
 O(a_1 - b_1, b_1)\cup O(a_1 + b_1, 0)&\textrm{  for}& a_1 > b_1 \\
 O(-a_1 + b_1, a_1) \cup O(a_1 + b_1, 0)&\textrm{  for }& a_1 < b_1
\end{array}\right.\;.
\end{align*}

\item
 When $0\neq a_1=b_1\neq 0$ and $a_2=b_2=0$,
  then
\begin{align*}
 &O(a_1,0)\otimes O(b_1,0)= \tfrac12O(2 a_1, 0)\cup \tfrac12 O(0, a_1) \cup\tfrac12O(2 a_1, 0)\cup \tfrac12 O(0, a_1)
 \\
  &\qquad\qquad\cup\tfrac12O( 0,a_1)\cup \tfrac12 O(0, a_1) =  O(2 a_1, 0)\cup2 O(0, a_1)\;.
 \end{align*}

\item
By analogy to above one can check that for $a_1,b_2\neq 0$ and $a_2=b_1=0$ the formula \eqref{A2product} takes the form
\begin{eqnarray*}
O(a_1,0)\otimes O(0,b_2)=\left\{
\begin{array}{lll}
O(a_1, b_2) \cup O(a_1 - b_2, 0)&\textrm{  for}& a_1 > b_2 \\
O(0, -a_1 + b_2)\cup O(a_1, b_2)&\textrm{  for }& a_1 < b_2\\
O(  a_1,a_1)\cup 3 O(0,0)&\textrm{  for }&a_1=b_2
\end{array}\right.\;.
\end{eqnarray*}

\item
The case  $O(0,a_2)\otimes O(0,b_2)$ is obtained by interchanging the first and second coordinates in each of orbits (when one wants to check  other propositions, one has to uses symmetries for appropriate groups or check such case separately).

\item
When one considers $a_1,a_2,b_1\neq 0$ all  different from each other and  $b_2=0$, then the product $O(a_1,a_2)\otimes O(b_1,0)$ could be simplified to the form
\begin{align*}
  & O(a_1,a_2)\otimes O(b_1,0)
  \\
  &\quad= O(a_1+b_1,a_2)\cup O(|a_1+\min\{0, a_2- b_1\}|,\min\{a_1,-a_2+ b_1\})
  \\
  &\qquad\qquad
\cup O(|a_1 - b_1|,a_2 + \min\{a_1, b_1\}).
  \end{align*}
Now all special cases should be considered separately, i.e.:
\begin{itemize}
  \item[$\circ$]
  for $a_2 >b_1  , a_1 > b_1$ one gets
\begin{align*}
  O(a_1,a_2)\otimes O(b_1,0&=O(a_1 + b_1, a_2)\cup O(a_1 - b_1, a_2 + b_1)
  \\
   &\qquad\qquad
  \cup O(a_1, a_2 - b_1)\;;
\end{align*}

  \item[$\circ$]
  for $a_2> b_1  , a_1 < b_1$ one gets
\begin{align*}
   O(a_1,a_2)\otimes O(b_1,0)&=O(a_1 + b_1, a_2)\cup O(-a_1 + b_1, a_1 + a_2)
   \\
   &\qquad\qquad\cup O(a_1, a_2 - b_1)\;;
\end{align*}

  \item[$\circ$]
  for $ a_2< b_1 , a_1 > b_1$ one gets
\begin{align*}
   O(a_1,a_2)\otimes O(b_1,0)&= O(a_1 + b_1, a_2)\cup O(a_1 - b_1, a_2 + b_1)
   \\
   &\qquad\qquad\cup O(a_1 + a_2 - b_1, -a_2 + b_1)\;;
\end{align*}

  \item[$\circ$]
  for $ a_2< b_1,a_1< b_1$ and $  a_1 + a_2 > b_1$ one gets
\begin{align*}
  O(a_1,a_2)\otimes O(b_1,0)&=O(a_1+b_1,a_2)\cup O(-a_1+b_1,a_1+a_2)
  \\
   &\qquad\qquad\cup O(a_1+a_2-b_1,-a_2+b_1)\;;
\end{align*}

  \item[$\circ$]
  for $ a_2 <b_1,a_1< b_1$ and $ a_1+a_2 <b_1$ one gets
\begin{align*}
  O(a_1,a_2)\otimes O(b_1,0)&=O(a_1 + b_1, a_2)\cup O(-a_1 + b_1,a_1 + a_2)
  \\
   &\qquad\qquad\cup O(-a_1-a_2+b_1,a_1).
\end{align*}

\end{itemize}

\item
In the case $a_1=b_1\neq 0$ and $b_2=0$ one gets
\begin{align*}
&O(a_1, a_2)\otimes O( a_1, 0)=O(2a_1,a_2)\cup O(\min\{a_1,a_2\},|a_1-a_2|)
\\
&\quad=\left\{\begin{array}{lll}
O(2 a_1, a_2)\cup  2 O(0, a_1 + a_2)\cup O(a_1, -a_1 + a_2)& \textrm{ for }&a_1 < a_2\\
O(2 a_1, a_2)\cup 2 O(0, a_1 + a_2)\cup O(a_2, a_1 - a_2)& \textrm{ for }&a_1 > a_2
\end{array}\right. .
\end{align*}

\item
For other special case $a_1=a_2=b_1\neq 0$ and $b_2=0$ one gets  another subcases:
\begin{align*}
    O(a_1, a_1)\otimes O( a_1, 0)&=O(2 a_1, a_1)\cup  2 O(0, 2a_1 ) \cup O(a_1,0)\;.
\end{align*}

\item
Because of the biggest amount of subcases the most difficult to verify  is the generic case, when   $a_1,a_2,b_1,b_2\neq 0.$ Below some  of   them are presented:
\begin{itemize}
    \item[$\circ$]
    for $-a_1 + b_2 > a_2 - b_1 , a_1 < b_1 $ and $   a_2 > b_2 , a_1 + a_2 >b_1 ,  a_1 > b_2$
\begin{align*}
 &\!\!\!\!\!\!\!\!\!\!\!\! O(a_1,a_2)\otimes O(b_1,b_2)
\\
&\!\!\!\!  =O(a_1 + b_1, a_2 + b_2) \cup O(a_1 + a_2 - b_1, -a_2 + b_1 + b_2)
 \\
 & \!\!\!\! \cup O(-a_1 + b_1, a_1 + a_2 + b_2) \cup O(-a_1 + b_1 + b_2, a_1 + a_2 - b_2)
     \\
 & \!\!\!\! \cup O(a_1 + b_1 + b_2, a_2 - b_2)\cup O(-a_1 - a_2 + b_1 + b_2, a_1 - b_2)\;;
\end{align*}

    \item[$\circ$]
    for $ -a_1 + b_2 > a_2 - b_1$ and $ a_1 < b_1, a_2 > b_2 $ and $  a_1 + a_2 > b_1 $ and $  a_1 < b_2$
\begin{align*}
&\!\!\!\!\!\!\!\!\!\!\!\! O(a_1,a_2)\otimes O(b_1,b_2)
\\
&\!\!\!\!= O(a_1 + b_1, a_2 + b_2)\cup   O(a_1 + a_2 - b_1, -a_2 + b_1 + b_2)
     \\
&\!\!\!\!  \cup O( -a_1 + b_1, a_1 + a_2 + b_2)
\cup O(-a_1 + b_1 + b_2, a_1 + a_2 - b_2)
\\
&\!\!\!\!\cup O(|a_2 - b_1|, -a_1 - a_2 + b_1 + b_2 + \min\{0, a_2 - b_1\})\;;
\\
& \!\!\!\!\cup \!O(a_1 + b_1 + b_2, a_2 - b_2)
\end{align*}

    \item[$\circ$]
    for $-a_1 + b_2 > a_2 - b_1 ,a_1 < b_1$ and $ a_2 > b_2$  and  $ a_1 + a_2 <b_1$
\begin{align*}
&\!\!\!\!\!\!\!\!\!\!\!\! O(a_1,a_2)\otimes O(b_1,b_2)=O(a_1 + b_1, a_2 + b_2)\cup O(-a_1 + b_1, a_1 + a_2 + b_2)
  \\
& \!\!\!\! \cup O(-a_1 - a_2 + b_1, a_1 + b_2)
\cup O(-a_1 + b_1 + b_2, a_1 + a_2 - b_2)
  \\
& \!\!\!\!\cup O(a_1 + b_1 + b_2, a_2 - b_2)
\\
&\!\!\!\!\cup O(|a_1 + a_2 - b_1 - b_2 + |\min\{0, a_1 - b_2\}||, |a_1 - b_2|)\;;
 \end{align*}

    \item[$\circ$]
    for $-a_1 + b_2 > a_2 - b_1 $ and $ a_1 < b_1 ,a_2 < b_2$
\begin{align*}
&\!\!\!\!\!\!\!\!O(a_1,a_2)\otimes O(b_1,b_2)=O(a_1 + b_1, a_2 + b_2)\cup O(-a_1 + b_1, a_1 + a_2 + b_2)
\\
&\!\!\!\! \cup O(a_1 + a_2 + b_1, -a_2 + b_2)
 \\
&\!\!\!\!\cup O(b_1 + \min\{a_2, -a_1 + b_2\}, |a_1 + a_2 - b_2|)
\\
& \!\!\!\!
\cup O( |a_1 + a_2 - b_1|, b_2 + \min\{a_1, -a_2 + b_1\})
\\
& \!\!\!\!
\cup O\left(|a_1\! + a_2\! - b_1\! - b_2 + |\!\min\{0, a_1\! - b_2\}|  |,\right.
 \\
&\qquad\qquad\left.
|a_1\! +     a_2\! - b_1 \!- b_2\! + |\!\min\{0, a_2\! - b_1\}| |\right)\;;
\end{align*}

    \item[$\circ$]
    for $   -a_1 + b_2 > a_2 - b_1 , a_1 > b_1$
\begin{align*}
&\!\!\!\!\!\!\!\!O(a_1,a_2)\otimes O(b_1,b_2)=O(a_1 + b_1, a_2 + b_2)\cup O(a_1 - b_1, a_2 + b_1 + b_2)
 \\
 & \!\!\!\!
\cup O(a_1 + a_2 - b_1, -a_2 + b_1 + b_2)
\cup O(a_1 + a_2 + b_1, -a_2 + b_2)
 \\
 & \!\!\!\!
\cup O(b_1 + \min\{a_2, -a_1 + b_2\}, |a_1 + a_2 - b_2|)
 \\
 & \!\!\!\!
\cup O\left(|a_1\! + a_2\! - b_1\! - b_2 +|\!\min\{0, a_1\! - b_2\}||\right.
\\
&\qquad\qquad\left., |a_1\! +
     a_2\! - b_1\! - b_2 \!+|\! \min\{0, a_2 \!- b_1\}| |\right)\;;
\end{align*}

    \item[$\circ$]
 for $ -a_1 + b_2 < a_2 - b_1 , a_2 > b_1$
\begin{align*}
&\!\!\!\!\!\!\!\!O(a_1,a_2)\otimes O(b_1,b_2)=O(a_1 + b_1, a_2 + b_2)
\\
& \!\!\!\!
\cup O(|a_1 - b_1|, a_2 + b_2 + \min\{a_1, b_1\})
\\
&\!\!\!\!\cup O(a_1 + b_1 + \min\{a_2, b_2\}, |a_2 - b_2|)
\\
&\!\!\!\! \cup O(|-a_1 + b_1 + b_2|, a_2 + \min\{b_1, a_1 - b_2\})
\\
& \!\!\!\!
\cup O(a_1 + \min\{a_2 - b_1, b_2\}, |-a_2 + b_1 + b_2|)
\\
& \!\!\!\!
\cup O(|-a_1 + b_2|, |-a_1 - a_2 + b_1 + b_2 + |\min\{0, -a_1 + b_2\}|| ).
\end{align*}

\end{itemize}

\item
In the end one could present some special cases
\begin{align*}
&\quad O(a_1,a_1)\otimes O(a_1,a_1)=
\\
&\qquad\qquad O(2 a_1, 2 a_1)\cup  2 O(0, 3 a_1)\cup 2 O(a_1, a_1)\cup  2O(3 a_1, 0)\cup 6 O(0, 0)\;;
\\
&\quad O(a_1,a_2)\otimes O(a_2,a_1)= O(a_1 + a_2, a_1 + a_2)
\\
&\qquad\qquad\cup O(a_1, a_1)\cup O(a_2, a_2)\cup O(|a_1 - a_2|, a_1 + a_2 + \min\{a_1, a_2\})
\\
&\qquad \qquad \qquad \cup O(a_1 + a_2 + \min\{a1, a2\}, |a_1 - a_2|)\cup 6 O(0, 0)\;.
\end{align*}\end{itemize}
One can check the rest of degenerated cases which are not describe to check formulae to the end. Each of them are easily verify directly. Hence the prove is completed.
\end{pf}
%:):):):):):):):):):):):):):):):):):):):):):):):):):):):):):):):):):):):):):):):):):):):):):):):):):):):):):):):)

\section{}

Below are presented  some special cases for $a,b>0$ and $a\neq b$:
\begin{align*}
O(a,a)\otimes O(a,a)&= O(2 a, 2 a)\cup 2 O(2 a, a)\cup 2 O(0, a)\cup   2 O(0, 3 a)
\\
&\qquad\cup 2 O(2 a, 0)\cup 2 O(4 a, 0)\cup 8 O(0, 0)\;;\\
O(a,b)\otimes O(a,b)&=O(2a,2b)\cup 2O(0,a)\cup 2O(0,a+2b)\cup 2O(2b,a)
\\
&\cup 2O(2a+2b,0)\cup 2O(2b,0)\cup 8O(0,0)\;;\\
 O(a,0)\otimes O(b,0)&=O(a+b,0)\cup O(|a-b|,0)\cup  O(|a-b|,\min\{a,b\})\;;\\
 O(a,b)\otimes O(b,a)&=O(a+b,a+b)\cup  O(a+b,a)\cup  O(a+b,b)
\\
&\!\!\!\!\!\!\!\!\!\!\!\!\cup  O(|a-b|,\min\{a,b\})\cup  O(|a-b|,a+b+\min\{a,b\})
\\
&\!\!\!\!\!\!\!\!\!\!\!\!\cup  O(a+b+\min\{a,b\},|a-b|)
\cup  O(a+b+2\min\{a,b\},|a-b|)
\\
&\!\!\!\!\!\!\!\!\!\!\!\!\cup  2O(|a-b|,0)\cup 2O(a+b,0)\;.
\end{align*}

\section{}

Then using formula \eqref{G2product} one could present special cases:
\begin{align*}
&O(a,a)\otimes O(a,a)= O(2 a, 2 a)\cup  2 O(a, 3 a)\cup 2 O(a, a) \cup 2 O(0, 5 a) \\
&\qquad\qquad \cup 2 O(0, 4 a)\cup 2 O(3 a, 0)\cup  2 O(2 a, 0)\cup 2 O(0, a)\cup 2 O(a, 0)\cup 12 O(0, 0)\;;
\\
&O(a,0)\otimes O(a,0)=O(2a,0)\cup 2O(a,0)\cup 2O(0,3a)\cup6 O(0,0)\;;\\
&O(a,b)\otimes O(a,b)=O(2a,2b)\cup 2O(0,b)\cup 2O(0,3a+b)\cup 2O(0,3a+2b)
\\&\qquad\qquad
\cup 2O(a,0)\cup 2O(a,b)\cup 2O(b,3a)\cup 2O(a+b,0)
\\&\qquad\qquad \qquad\qquad \cup 2O(2a+b,0)\cup 12O(0,0)\;;\\
 &O(a,b)\otimes O(b,a)=O(a+b,a+b)\cup O(a,a+2b) \cup O(b,2a+b)
 \\
 &\qquad\qquad
\cup O(|a-b|,a+b+\min\{3a,3b\}) \cup O(\min\{a,2b\},|a-2b|)
 \\
 &\qquad\qquad
 \cup O(a+b+\min\{a,b\},|a-b|)\cup O(\min\{a,b\},|a-b|)
 \\
 &\qquad\qquad\qquad\cup O(\min\{2a,b\},|b-2a|)\cup 2O(0,2a+2b)\cup 2O(0,|a-b|).
\end{align*}

\section{}

Non-crystallographic group $H_2$ differ from crystallographic ones which are described in many papers. Below  some facts concerning $H_2$, collected from  \cite{ H, HLP08, P95,ChMP98, T01, PT02} are presented.

\begin{figure}[h!]
\begin{center}
{\includegraphics[scale=1.6]{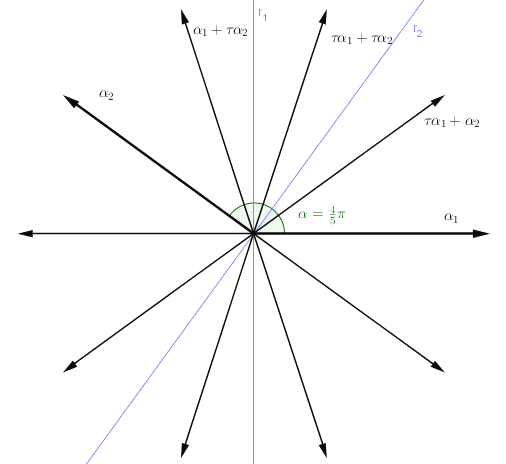}}
\caption{The roots of group $H_2$ and reflections $r_1,r_2$}\label{pierwiastkiH2}
\end{center}
\end{figure}

In the complex plane root system for $H_2$ could be  a set of $10$th roots of unity
$$
\left\{\pm\zeta^j\mid\zeta=e^{\tfrac{2\pi i}{5}}\right\}\subset \C.
$$
In this paper   the simple roots were chosen as  $\a_1=\sqrt2\zeta,$ $\a_2=\sqrt2\zeta^2,$ then the highest root is $\xi=\tau\a_1+\tau\a_2$ (see figure \ref{pierwiastkiH2}) and
$$
\vartriangle=\left\{ \pm\a_1,\pm\a_2,\pm(\a_1+\tau\a_2),\pm(\tau\a_1+\a_2),\pm\tau(\a_1+\a_2)\right\},
$$
where $
\tau=\tfrac12(1+\sqrt5)=2\cos \tfrac\pi5,$ $\tau'=\tfrac12(1-\sqrt5)=-2\cos \tfrac{2\pi}{5}.
$  There are solutions of the equation on golden ratio $x^2-x-1=0$. It is easy to check that $\tau\cdot\tau'=-1,$   $\tau+\tau'=1.$% and $\tau^2=\tau+1.$
\

\begin{figure}[h!]
\begin{center}
{\includegraphics[scale=0.35]{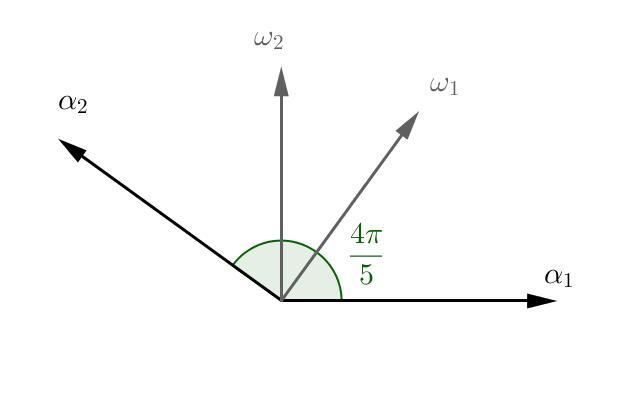}}
\caption{Dual basis to $\a$ and $\beta$}\label{dualbasis}
\end{center}
\end{figure}

The Cartan matrix of $H_2$ and inverse Cartan matrix are:
$$
C=\left(\frac{2\langle\alpha_i\mid\a_j\rangle}{\langle\alpha_j\mid\alpha_j\rangle}\right)=
\left(
  \begin{array}{cc}
    2 & -\tau \\
    -\tau & 2 \\
  \end{array}
\right),\qquad
C^{-1}=\tfrac{1}{5}
\left(
  \begin{array}{cc}
    4+2\tau & 1+3\tau \\
    1+3\tau & 4+2\tau \\
  \end{array}
\right).
$$
The relation between  the dual basis is given in terms of Cartan matrix, see \eqref{basesH2} and it is presented in figure \ref{dualbasis}.

 \

 After presentation of basic facts for the group $H_2$ one could  calculate some special cases of product of two orbits for $a,b>0$ and $a\neq b$
\begin{align*}
O(a,a)\otimes O(a,a)&=  O(2 a, 2 a)\cup  2 O(a\tau, a\tau)\cup  2 O(-a+a\tau ,-a+a\tau )
\\
&\cup 2 O(0,-a+2a\tau ) \cup 2 O(-a+2a\tau, 0)
\\
&\qquad
\cup
 2 O(2a+a\tau, 0)\cup 2 O(0,2a+a\tau)\cup 10 O(0, 0)  \;;\\
  O(a,0)\otimes O(0,a)&=O(a,a)\cup O(-a+a\tau,-a+a\tau)\cup5O(0,0)\;;\\
    O(a,b)\otimes O(a,b)&=O(2a,2b)\cup 2O(b\tau,a\tau)\cup2O(-a+a\tau,-b+b\tau)\\
  & \cup2O(0,2b+a\tau)\cup 2O(2a+b\tau,0)\\
  &\quad
  \cup 2O(0,-b+\tau(a+b))\cup2O(-a+\tau(a+b),0)\\
 &\quad\quad\cup2O(\max\{0,a-b\}(-1+\tau),\max\{0,b-a\}(-1+\tau))\;;\\
  O(a,b)\otimes O(b,a)&
\\
&\!\!\!\!\!\!\!\!\!\!\!\!\!\!\!\!\!\!\!\!\!\!\!\!\!\!\!\!\!\!\!\!\!\!\!\!\!\!\!\!=O(a+b,a+b)\cup O(b-a+a\tau,b-a+a\tau)  \cup O(a-b+b\tau,a-b+b\tau)\\
 & \!\!\!\!\!\!\!\!\!\!\!\!\!\!\!\!\!\!\!\!\!\!\!\!\!\!\!\!\!\!\!\! \cup O(|a-b|,a+b+\min\{a,b\}\tau) \cup O(a+b+\min\{a,b\}\tau,|a-b|)\\
 &\!\!\!\!\!\!\!\!\!\!\!\!\!\!\!\!\!\!\!\!\!\!\!\!\!\!\!\!\!\!\!\!\cup O((1-\tau)(\min\{a,b\}-\max\{a,b\}),\min\{|a - b \tau - a \tau |, |b - a \tau - b \tau|\})
 \\
 &\!\!\!\!\!\!\!\!\!\!\!\!\!\!\!\!\!\!\!\!\!\!\!\!\!\!\!\!\!\!\!\!\cup O(\min\{|a - b \tau - a \tau |, |b - a \tau - b \tau|\},(1-\tau)(\min\{a,b\}-\max\{a,b\}))\\
  &\quad \cup O(-b+b\tau,-b+b\tau) \cup O(-a+a\tau,-a+a\tau)\cup 10 O(0,0)\;.
\end{align*}
Orbits products for $H_2$ are the most interesting. Because of properties of $\tau$ it could be rewrite  in very different form, for example
\begin{align*}
O(a,0)\otimes O(a,0)&=O(2a,0)\cup2O(0,a\tau)\cup2O( \tfrac{a}{\tau},0)
\\
&=O(2a,0)\cup2O(0,a\tau)\cup2O(-a+a\tau,0)\;
\end{align*}
and also one can calculate orbit products  for multiplication of $\tau$ or $-\tau'=-1+\tau=\tfrac{1}{\tau}$
\begin{align*}
O(\tau,0)\otimes O(\tau,0)&=O(2\tau,0)\cup 2O(0,1+\tau)\cup 2O(1,0)\;;\\
O(\tau,0)\otimes O(\tfrac1\tau,0)&=O(\tau,0)\otimes O(\tau-1,0)
\\&=O(1,1)\cup O(2\tau-1,0)\cup O(\tau-1,\tau-1)\\
&=O(1,1)\cup O(\tfrac{2+\tau}{\tau},0)\cup O(\tfrac1\tau,\tfrac1\tau)\;;\\
O(\tfrac{1}{\tau},0)\otimes O(\tfrac{1}{\tau},0)&=O(-1+\tau,0)\otimes O(-1+\tau,0)=O(-\tau',0)\otimes O(-\tau',0)\\
&=O(\tfrac{2}{\tau},0)\cup2O(\tfrac{1}{1+\tau},0)\cup2O(0,1)\\
&=O(-2+2\tau,0)\cup2O(2-\tau,0)\cup2O(0,1)\\
&=O(-2\tau',0)\cup2O(1+\tau',0)\cup2O(0,1).
\end{align*}
This group has the largest potential among all described groups in this paper and it is a very good staring point to another paper.
%%%%%%%%%%%%%%%%%%%%%%%%%%%%%%%%%%%%%%%%%%%%%%%%%%%%%%%%%%%

%%%%%%%%%%%%%%%%%%%%%%%%%%%%%%%%%%%%%%%%%%%%%%%%%%%%%%%%%%%
%%%%%%%%%%%%%%%%%%%%%%%%%%%%%%%%%%%%%%%%%%%%%%%%%%%%%%

%%%%%%%%%%%%%%%%%%%%%%%%%%%%%%%%%%%%%%%%%%%%%%%%%%%%%%%%%%%
%\section*{References}

\end{document}